\documentclass[11pt]{article}
\pdfoutput=1
\usepackage{jheppub}

\usepackage{tikz}
\usepackage{amsfonts,amsmath,amssymb,amsthm}
\usepackage{mathrsfs,bm,bbm,esint}
\usepackage[mathscr]{eucal}
\usepackage{physics}
\usepackage{slashed,soul}
\usepackage{rotating,pdflscape}
\usepackage{enumitem}
\usepackage{multirow,array}
\usepackage[numbers,sort&compress]{natbib}
\usepackage{textgreek}

\newcommand{\fixM}{\text{\bf{\textgamma}}}

\title{Time-independence of gravitational R\'enyi entropies and unitarity in quantum gravity}

\author{Donald Marolf,}
\author{Zhencheng Wang}
\affiliation{
Department of Physics, University of California, Santa Barbara, CA 93106, USA}
\emailAdd{marolf@ucsb.edu}
\emailAdd{zhencheng@ucsb.edu}

\abstract{
The Hubeny-Rangamani-Takayanagi surface $\gamma_{HRT}$ computing the entropy $S(D)$ of a domain of dependence $D$ on an asymptotically AdS boundary is known to be causally inaccessible from $D$.     We generalize this gravitational result to higher replica numbers $n >1$ by considering the  replica-invariant surfaces (aka `splitting surfaces') $\text{\bf{\textgamma}}$  of
real-time replica-wormhole saddle-points computing R\'enyi entropies $S_n(D)$ and showing that there is a sense in which $D$ must again be causally inaccessible from $\text{\bf{\textgamma}}$ when the saddle preserves both replica and conjugation symmetry.  This property turns out to imply  the $S_n(D)$ to be independent of any choice of any Cauchy surface $\Sigma_D$ for $D$, and also that the $S_n(D)$ are independent of the choice of boundary sources within $D$. This is a key hallmark of  unitary evolution in any dual field theory.  Furthermore, from the bulk point of view it adds to the evidence that time evolution of asymptotic observables in quantum gravity is implemented by a unitary operator in each baby universe superselection sector.  Though we focus here on pure Einstein-Hilbert gravity and its Kaluza-Klein reductions,  we expect the argument to extend to any two-derivative theory who satisfies the null convergence condition.  We consider both classical saddles and the effect of back-reaction from quantum corrections.
}

\begin{document}
\maketitle


\section{Introduction}
\label{sec:intro}

Many points of view have long motivated the idea that,  in order to describe measurements of distant observers, black holes can be modeled as a quantum system with density
of states $e^{S_{BH}}$ whose evolution is unitary up to possible interactions with other
quantum systems; see e.g. \cite{Mathur:2009hf,Harlow:2014yka,Marolf:2017jkr} for reviews.  Here $S_{BH}$ is the Bekenstein-Hawking entropy of the black hole. Following \cite{Marolf:2020rpm}, we refer to the above idea as Bekenstein-Hawking unitarity (or BH unitarity).\footnote{It was instead called `the central dogma' in \cite{Almheiri:2020cfm} in analogy with the term's use in biology.}  This property would in particular imply that Hawking radiation from evaporating black holes must carry information in a manner famously described by Page \cite{Page:1993df}.  The recent replica wormhole derivations \cite{Almheiri:2019qdq,Penington:2019kki} of the expected `Page curve' for the entropy of this radiation
thus provide strong evidence that there is a sense in which BH unitarity holds.

Our goal here is to provide additional support for the idea that time evolution in quantum gravity is implemented by unitary operators on appropriate Hilbert spaces.  For simplicity, we consider asymptotically AdS spacetimes.  In that context one may think of this program as further verifying properties predicted by unitarity in some dual field theory description.  But one may also take point of view that the bulk theory decomposes into superselection sectors defined by states in a so-called `baby universe' sector of the theory, and that we verify predictions of the hypothesis that time evolution is unitary in each such superselection sector; see  \cite{Coleman:1988cy,Giddings:1988cx,Giddings:1988wv}, related remarks in \cite{Saad:2018bqo,Saad:2019lba,Penington:2019kki}, and axiomatic arguments in \cite{Marolf:2020xie,Marolf:2020rpm}.

In this context, recall that a key hallmark of unitarity is the preservation of the eigenvalues of any density matrix in an isolated system.\footnote{In particular,  this result (or even just preservation of von Neumann entropy) implies that time evolution maps pure states to pure states.  If one also assumes time evolution to be a quantum channel, then the channel must in fact be unitary for this property to hold. This conclusion follows from the so-called Stinespring dilation theorem \cite{Stinespring:1955}, which allows any quantum channel to be represented by tensoring the given system with some ancilla state, acting with a unitary on the joint system, and then tracing out the ancilla.  If this procedure maps pure states to pure states, then the joint-system unitary cannot create entanglement with the ancilla and must thus define a unitary on the original system alone.  We thank Geoffrey Penington for discussions regarding this point.}  This can be readily probed, and in some cases proven, by checking time-independence of the associated R\'enyi entropies $S_n
 = - \frac{1}{n-1} \log\left(\frac{{\rm Tr}\left[\left(\rho(D)\right)^n\right]}{\left[{\rm Tr} \left(\rho(D)\right)\right]^n}\right)$.

In the context of Einstein-Hilbert gravitational systems, an analogous result was established in  \cite{Wall:2012uf,Headrick:2014cta}.  These works considered the
Hubeny-Rangamani-Takayanagi (HRT) surface $\gamma_{HRT}$ \cite{Hubeny:2007xt} associated with the entropy $S(D)$ of a domain of dependence $D$ on an asymptotically AdS boundary.  They then used the bulk Raychaudhuri equation to show that $\gamma_{HRT}$  must be causally inaccessible from $D$ when the Lorentzian bulk spacetime satisfies the null energy condition. As a result,
no choice of boundary conditions on $D$ can influence the HRT entropy determined by the area of $\gamma_{HRT}$.  In other words, the von Neumann entropy is invariant under any time evolution for which the system can be said to remain closed.   But corresponding results remain to be established for $S_n(D)$ with $n>1$.

This work begins to bridge this gap by studying saddle points of the real-time gravitational path integral for $S_n(D)$.
In the context of an AdS bulk that is dual to a unique field theory, this $S_n(D)$ is the standard R\'enyi entropy.  But in general the quantity
$S_n(D)$ computed by these path integrals is more accurately described as a so-called swap R\'enyi entropy\footnote{From the purely bulk perspective, the term `swap entropy' is physically most appropriate when we couple the AdS system to a non-gravitational bath and compute a (swap) entropy for some subset of the bath.  We will nevertheless also use it in the above context where no bath is present.} \cite{Marolf:2020rpm}; see also \cite{Giddings:2020yes}.

Indeed, for reasons that we now describe, it would be even better to call the $S_n(D)$ we study an `annealed swap R\'enyi entropy.'  In a baby universe scenario for the bulk, this $S_n(D)$ is expected to closely approximate the average over superselection sectors of the corresponding R\'enyi entropy in each sector \cite{Marolf:2020xie,Marolf:2020rpm}.  However, it will differ from this average for two reasons:  The first is that it is an `annealed average,' meaning that we actually study the average $\exp(-S_n)$ and then take a logarithm.  The second is a similar issue due to the fact that the normalization $Tr \rho$ can vary among members of the ensemble,  but we normalize only by the average of $Tr \rho$.  I.e., denoting averaging over superselection sectors by an overline, our gravitational Re\'nyi gives
\begin{equation}
\label{eq:Snav}
S_n
: = - \frac{1}{n-1} \log\left(\frac{\overline{{\rm Tr}\left[\left(\rho(D)\right)^n\right]}}{\left[\overline{{\rm Tr} \left(\rho(D)\right)}\right]^n}\right).
\end{equation}
Expression \eqref{eq:Snav} can also be used to relate our gravitational Re\'nyi to averages of ${\rm Tr}\left[\left(\rho(D)\right)^n\right]$ over any ensemble of dual theories.    In either context, we will use the term annealed swap Re\'nyi to refer to all of the complications of \eqref{eq:Snav}.

We consider contexts where the real-time (swap) replica path integral  is dominated by a saddle that preserves both replica and conjugation symmetry.  Examples of such real-time saddles were recently presented in \cite{Colin-Ellerin:2021jev}.  Such path integrals and their saddles were described in \cite{Dong:2016hjy,Marolf:2020rpm} and especially \cite{Colin-Ellerin:2020mva}. While parts of the saddle-point spacetime have a complex-valued metric, with the above symmetries there is a real Lorentz-signature metric on the regions spacelike separated from the replica-invariant surface $\fixM$ (aka `the splitting surface') \cite{Colin-Ellerin:2020mva}.  And while the real-time saddles described in
\cite{Colin-Ellerin:2020mva} are singular at $\fixM$,  we show that there is an appropriate sense in which $\fixM$ remains extremal for $n>1$.

For theories that satisfy the null energy condition on-shell, by making one assumption it will then follow from the results of \cite{Wall:2012uf,Headrick:2014cta} that $\fixM$ must again be causally inaccessible\footnote{Since only part of the spacetime is real and of Lorentz signature, this phrase remains to be properly defined.  It will be discussed briefly in section \ref{sec:summary} and in more detail in section \ref{sec:classical}. } from $D$.
Although the argument is more subtle than in the HRT context, this will then again imply the Renyi entropies $S_n(D)$ to be independent of any choices within $D$.  The key point is that our saddles will extend into the future only up to some surface $\Sigma_{M_-}$ that is spacelike separated from $\fixM$.  In particular, much as in a Schwinger-Keldysh contour, our saddles will contain both future-directed pieces of spacetime and past-directed pieces of spacetime that meet in a timefold on $\Sigma_{M_-}$.  The replica and conjugation symmetries require the solutions on future- and past-directed pieces to be related by complex conjugation, so that coincide in the region spacelike related to $\fixM$ where the solution is real.  Since the future- and past-directed pieces are weighted respectively by $e^{iS}$ and $e^{-iS}$ in the path integral, the contributions from the region spacelike separate from $\fixM$ cancel, and the result is unchanged if we simply take $\Sigma_{M_-}$ to lie along the past light cone of $\fixM$.  But doing so removes the entirety of $D$ from the boundary so that the remaining saddle is manifestly independent of choices within $D$.

Furthermore, the reader may recall from \cite{Engelhardt:2014gca} that one may use the generalized second law (GSL)
to upgrade the arguments of \cite{Wall:2012uf,Headrick:2014cta} to include quantum corrections.  The same will again be true for our $n>1$ R\'enyi problem.   (One may of course consider the the generalized second law to follow from the
quantum focussing condition of \cite{Bousso:2015mna}.)

Throughout this work we focus on the case of Einstein-Hilbert gravity with minimal couplings to any matter fields, though as noted in \cite{Colin-Ellerin:2020mva} the generalization to higher-derivative gravity is straightforward.
For the interested reader, a brief summary of the argument for extremality of $\fixM$ can be found in section \ref{sec:summary} below.  Other readers may prefer to proceed directly to the main discussion of sections \ref{sec:review}-\ref{sec:disc}.

The main text will begin with a brief review of the real-time gravitational path integral for $S_n(D)$ following \cite{Dong:2016hjy,Marolf:2020rpm} and especially \cite{Colin-Ellerin:2020mva}.  This material is presented in section \ref{sec:review}, along with a description of the relevant saddle points.  Section \ref{sec:classical} then argues at the classical level that the splitting surface $\fixM$ is extremal in saddles that preserve replica and conjugation symmetry.  It also discusses the precise sense in which this requires $\fixM$ to be causally inaccessible from $D$ and in which it makes $S_n(D)$ independent of sources on $D$ or choices of Cauchy surfaces $\Sigma_D$ for $D$.  Section \ref{sec:quantum} then follows with the upgraded argument that includes quantum corrections.   As a supplement to the quantum argument, appendix \ref{app:RSHRT} illustrates the stationarity of $S_{QFT}$ at $\fixM$  an example in which the quantum corrections come from a bulk QFT that happens to be holographic; i.e., in which the matter entropy is described by an HRT surface in a higher-dimensional spacetime.  We close with a broader discussion of unitarity in quantum gravity in section \ref{sec:disc}.

\subsection{Summary of the extremality arguments}
\label{sec:summary}

Let us briefly explain why $\fixM$ should be extremal in our R\'enyi problem.  As described in \cite{Colin-Ellerin:2020mva}, in a saddle point geometry ${\cal M}_n$ for the real-time $n$-replica path integral, the metric near the splitting surface must have an asymptotic expansion of a certain form. This expansion appears singular as presented in \cite{Colin-Ellerin:2020mva}, but as explained there it in fact matches the form that would be obtained by applying a particular Wick rotation to a smooth Euclidean space ${\cal M}_n^E$, and in particular to a Euclidean space with no conical singularities. It is merely that the Wick rotation makes use of singular coordinates on the smooth geometry ${\cal M}^E_n$.     Furthermore, the fact that the original real-time saddle ${\cal M}_n$ preserves replica symmetry requires ${\cal M}^E_n$ to have a $\mathbb{Z}_n$ symmetry that preserves the splitting surface $\fixM$, but that rotates the tangent space at each point of $\fixM$ in the plane orthogonal to $\fixM$.   Since the codimension-2 extrinsic curvature of $\fixM$ must be invariant under this $\mathbb{Z}_n$ rotation, it must in fact vanish.  In particular,  the trace of the extrinsic curvature vanishes and $\fixM$ is extremal in ${\cal M}^E_n$.

While the splitting surface $\fixM$ is necessarily singular in the original real-time solution, the above facts imply that there is an arbitrarily smooth\footnote{I.e., $\hat {\cal S}_R$ can be chosen to be $C^m$ for arbitrary $m$.} spacetime $\hat{\cal S}_R$ (the `right shadow of ${\cal M}_n$') with an extremal surface $\fixM$ such that $\hat{\cal S}_R$ coincides with any single sheet of the original real saddle ${\cal M}_n$ in a connected region spacelike separated from $\fixM$ that contains the boundary domain $D$.  Furthermore, $\hat{\cal S}_R$ can be constructed by finding `smoother' coordinates in the original real-time solution without reference to analytic continuation on ${\cal M}^E_n$.
We will refer to this region of ${\cal M}_n$ as the `right wedge\footnote{As an $n$-replica geometry, ${\cal M}_n$ in fact contains $n$ such regions, but we may choose any one to call the right wedge.}' of ${\cal M}_n$, which is why and we call $\hat{\cal S}_R$ the `right shadow.'  Although its geometry depends on $n$, we suppress the label $n$ on $\hat{\cal S}_R$.

Finally, as noted in \cite{Colin-Ellerin:2020mva} the above symmetries require the metric to be real and Lorentz signature in the region spacelike separated from $\fixM$.  As a result, we may take  $\hat{\cal S}_R$ to be both real and Lorentz-signature; indeed, this is what gives a well defined notion of `the region spacelike separated from $\fixM$' used above. In addition, since the null convergence condition holds where $\hat{\cal S}_R$ coincides with the original saddle ${\cal M}_n$, by taking limits it also holds on the closure of this region in  $\hat{\cal S}_R$.  But in real Lorentz-signature spacetimes satisfying the null convergence condition, refs. \cite{Wall:2012uf,Headrick:2014cta}  showed any extremal surface anchored to the boundaries of Cauchy surfaces of $D$ to be causally inacessible from $D$.  The fact that $\hat{\cal S}_R$  and ${\cal M}_n$ coincide in the region spacelike separated from $\fixM$ then provides a sense in which this conclusion also holds in ${\cal M}_n$; see section \ref{sec:classical} for details.

The above perspective on the classical case now suggests a generalization that includes quantum corrections.  If the splitting surface is also a stationary point of the generalized entropy $S_{gen}[\gamma] = \frac{A}{4G} + S_{QFT}$, then the quantum focussing condition of \cite{Bousso:2015mna} (or, indeed, just the GSL as in \cite{Engelhardt:2014gca}) will again allow us to conclude that $\fixM$ must be causally separated from $D$ \cite{Engelhardt:2014gca}.

One would thus like to argue as in the classical case that replica symmetry requires $S_{gen}$ to be stationary on $\fixM$.  However, for any codimension-2 surface $\gamma$ (which will generally differ from $\fixM$), the bulk entropy $S_{QFT}[\gamma]$ refers to the entropy of quantum fields on a partial Cauchy surface stretching from $\gamma$ to some boundary region.  As shown in figure \ref{fig:Sbulk}, this choice manifestly breaks replica symmetry and thus appears to invalidate the desired argument.  Indeed, as shown in the same figure, the very requirement that $\gamma$ can be connected to this boundary region by a partial Cauchy surface means if $S_{QFT}[\gamma]$  is defined at all, then $S_{QFT}[\gamma']$ will {\it not} be defined for any $\gamma'$ related to $\gamma$ by a non-trivial replica symmetry.

However, within the region where it is defined, $S_{QFT}$ can be computed by considering the response of the partition function for bulk quantum fields to a change in boundary conditions that {\it does} respect replica symmetry.\footnote{This is just the usual replica trick applied to quantum fields propagating on the real-time replica wormhole geometry; see section \ref{sec:quantum}.} As a result, one may extend the definition of $S_{QFT}$ in a manner that preserves replica symmetry.   We use this observation below to show that replica symmetry requires  both $A$ and $S_{QFT}$ to be separately stationary on the splitting surface $\fixM$, so that $S_{gen} = A/4G + S_{QFT}$ is stationary as well.

\section{Real-time path integrals with splitting surfaces}
\label{sec:review}

This section provides an extremely brief summary of the real-time gravitational path integral computation for $S_n(D)$ and the relevant saddles following \cite{Dong:2016hjy,Marolf:2020rpm} and especially \cite{Colin-Ellerin:2020mva}.      The general structure of such path integrals is described in section \ref{sec:gen}, while boundary conditions at a special `splitting surface' are described in section \ref{sec:split}. Saddles are further discussed in section \ref{sec:saddles}.  The reader may also wish to consult \cite{Colin-Ellerin:2021jev} for concrete examples.

\subsection{Real-time R\'enyi path integrals}
\label{sec:gen}

The boundary conditions for a gravitational path integral are typically chosen by first considering a corresponding non-gravitational problem.  This long tradition is often justified by appealing to AdS/CFT or a broader holographic principle, but as described in \cite{Marolf:2020rpm}, in many cases it also follows from a certain operational perspective associated  with coupling non-gravitational systems to the gravitational system of interest.\footnote{Though this may then be associated with refinements in the interpretation, such as replacing entropies by so-called swap entropies.}  But any of these perspectives motivates us to begin with a discussion of real-time R\'enyi path integrals for systems in which gravity is not dynamical.

To be specific, let us first consider the
real-time path integral computation of $S_n(D)$ in a non-gravitating relativistic quantum field theory on some Lorentz signature spacetime ${\cal B}$.   Recall that, given a (perhaps improperly normalized) density matrix $\rho(D)$ on a domain of dependence $D$, we define
\begin{equation}
S_n(D) : = - \frac{1}{n-1} \log\left(\frac{{\rm Tr}\left[\left(\rho(D)\right)^n\right]}{\left[{\rm Tr} \left(\rho(D)\right)\right]^n}\right).
\end{equation}

Furthermore, given a pure state $|\psi\rangle$ on the entire system, one can define the density matrix $\rho(D)$ associated with some domain of dependence $D \subset {\cal B}$ by tracing $|\psi\rangle \langle \psi |$ over some Cauchy surface $\Sigma_{\bar D}$ for the region $\bar D \subset {\cal B}$ that is spacelike separated from $D$.  Powers $(\rho(D))^n$ can then be computed by evaluating each copy of $\rho(D)$ on some cauchy surface $\Sigma_D$ of $D$ and performing appropriate index contractions. And we may similarly compute $\Tr [(\rho(D))^n]$ by contracting the final two free indices.   Unitarity of the QFT then requires the result to be independent of the choices of $\Sigma_{\bar D}$ and $\Sigma_D$.  It also requires the result to be unchanged by the addition of any sources on either $D$ or $\bar D$.

As a result, if we are given a representation of
$|\psi\rangle$ as a path integral over some manifold ${\cal B}_-$ with boundary $\partial {\cal B}_-$ containing $\Sigma_{\bar D} \cup \Sigma_D$,  it is straightforward to construct a path integral for $\Tr [(\rho(D))^n]$. In general, the manifold ${\cal B}_-$ might be Euclidean, Lorentzian, complex, or a Schwinger-Keldysh-like combination of these options.  However, the notation ${\cal B}_-$ reflects the fact that in many cases one would like to prepare some state in the past and then evolve it forward in time to $\Sigma_{\bar D} \cup \Sigma_D$.  In such cases ${\cal B}_-$ will contain at least the (real and Lorentz-signature) region of ${\cal B}_-$ immediately to the past of $\Sigma_{\bar D} \cup \Sigma_D$.

\begin{figure}
\centering
\includegraphics[scale=0.8]{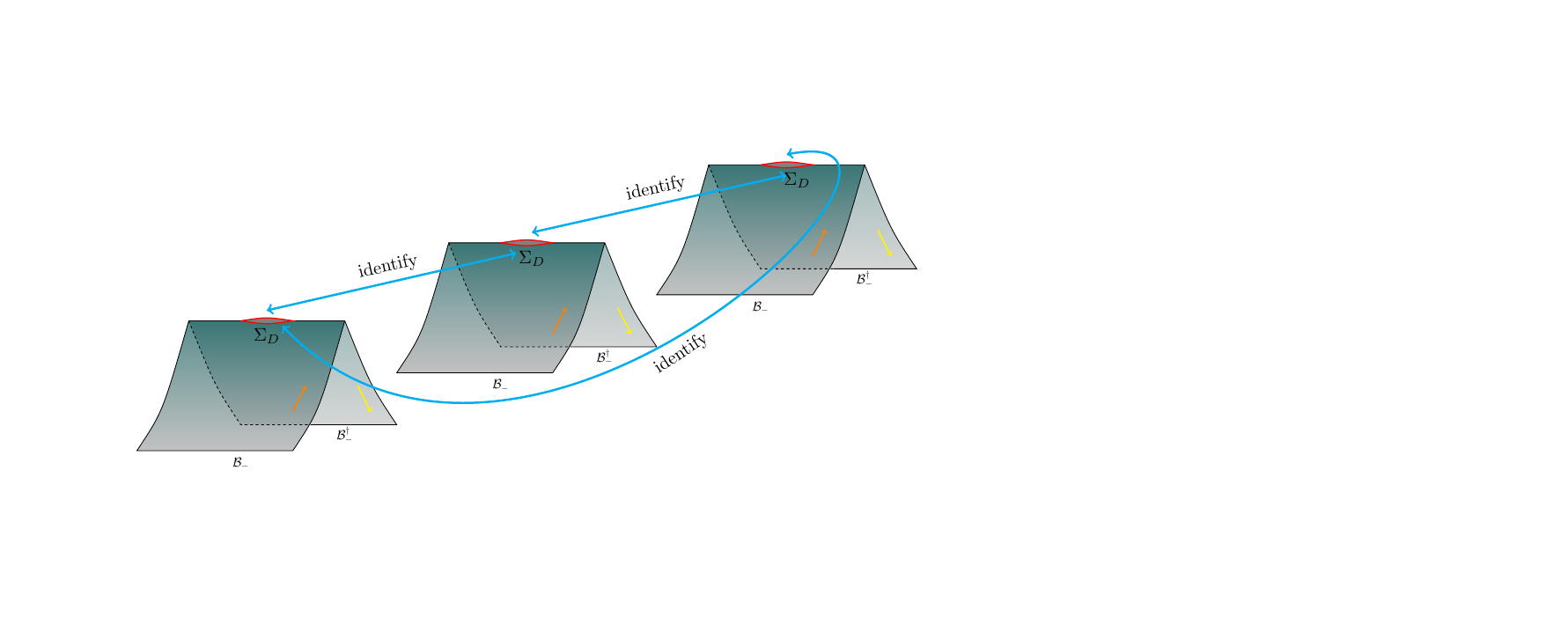}
\caption{The manifold $\mathcal{B}_n$ used  compute the path integral for $\Tr [(\rho(D))^n]$ in a non-gravitating QFT.  Here $\Sigma_D$ is a Cauchy surface of $D$ and the figure shows $n=3$.}
\label{fig:Bn}
\end{figure}

As shown in figure \ref{fig:Bn}, the path integral for $\Tr [(\rho(D))^n]$  is thus naturally represented as a path integral over a spacetime ${\cal B}_n$ constructed by cutting open $n$ copies of ${\cal B}_-$ and $n$ copies of an appropriate adjoint manifold\footnote{The path integral weight for a region of in the adjoint ${\cal B}^\dagger_-$ is the complex conjugate that for the corresponding region of ${\cal B}_-$.  In particular, sources in ${\cal B}^\dagger_-$ are the complex conjugates of sources in ${\cal B}_-$, and in terms of the Lorentz-signature action $S_L$ regions of ${\cal B}^\dagger_-$ are weighted by $e^{-iS_L}$ instead of $e^{iS_L}$.} ${\cal B}^\dagger_-$ and pasting them together in a manner that -- at least in simple cases -- results in ${\cal B}_n$ having only a single connected component.  However, even when ${\cal B}_-$ is real and Lorentz-signature near $\Sigma_{\bar D} \cup \Sigma_D$, the R\'enyi manifold ${\cal B}_n$ fails to have a standard causal structure on $\Sigma_{\bar D} \cup \Sigma_D$, and in particular at $\partial \Sigma_{\bar D} = \partial \Sigma_D$.  But the path integral remains well-defined, since on these surfaces it merely implements the above contractions.

We now turn to the corresponding gravitational problem in which one defines  R\'enyis via the standard formula
\begin{equation}\label{eq:renyiMn}
S_n(D) = \frac{1}{1-n}\, \log\left(\frac{Z[{\cal B}_n]}{Z[{\cal B}]^n}\right)  = \frac{1}{n-1} \left(I_n - n \, I_1 \right)\, ,
\end{equation}	
where $Z[{\cal B}_n]$  is the gravitational path integral over spacetimes with a fixed asymptotically-AdS boundary ${\cal B}_n$ of the form constructed above. In particular, ${\cal B}_n$  has a $\mathbb{Z}_n$ replica symmetry that cyclicly permutes copies of ${\cal B}_- \cup {\cal B}^\dagger_-$, as well as a conjugation symmetry that exchanges each copy of
${\cal B}_-$ with an associated ${\cal B}^\dagger_-$ (and thus complex-conjugating all sources).  However, the individual bulk configurations that contribute to
$Z[{\cal B}_n]$ are allowed to break either or both symmetries.  In the semiclassical limit, we can evaluate the above R\'enyis using
$I_n :=  -\log Z[{\cal B}_n]  \approx- i\, S_L [{\cal M}_n]$ for the appropriate saddles ${\cal M}_n$.

Now, the bulk spacetimes over which we integrate must be compatible with the structure of ${\cal B}_n$ near $\Sigma_{\bar D} \cup \Sigma_D$.  Following \cite{Marolf:2020rpm,Colin-Ellerin:2020mva} we specify the desired bulk spacetimes by first considering the set ${\sf M}_-$ of bulk spacetimes $M_-$ that contribute to the gravitational path integral with asymptotic boundary conditions ${\cal B}_-$, and thus which also end on some bulk surface $\Sigma_{M_-}$ with boundary $\partial \Sigma_M = \Sigma_{\bar D} \cup \Sigma_D$ at which the induced metric is to be fixed as an additional boundary condition.  The path integral $Z[{\cal B}_n]$ will then be defined to sum over bulk spacetimes ${\cal M}_n$ constructed by choosing $2n$ of the above spacetimes  $M^i_-, \tilde M^i_- \in {\sf M}_-$ for $i=1,\dots n$ and applying a cut-and-paste procedure to $M^i_-$ and the adjoint manifolds $\tilde M^{\dagger i}_-$ analogous to that described for ${\cal B}_n$ above.  Here, however, for each final bulk surface $\Sigma_{M^i_-},\Sigma_{\tilde M^{\dagger i}_-}$ we may choose an arbitrary\footnote{In particular, such splitting surfaces may have multiple connected components.  This allows for the R\'enyi equivalent of `islands' \cite{Almheiri:2019hni}.} `splitting surface' $\fixM_i, \tilde \fixM_i$ that partitions $\Sigma_{M^i_-},\Sigma_{\tilde M^{\dagger i}_-}$ into two pieces (in analogy with the way that $\partial {\Sigma_D}$ partitions $\Sigma_{\bar D} \cup \Sigma_D$ into $\Sigma_{\bar D}$ and $\Sigma_D$).  Furthermore, since the bulk manifolds $\Sigma_{M^i_-},\Sigma_{\tilde M^{\dagger i}_-}$ are generally distinct, the bulk metric generally fails to be continuous where two such surfaces are pasted together.  When ${\cal M}_n$  is discontinuous in this way, we take it to have zero amplitude; i.e, the process of pasting together such surfaces leads to an appropriate delta-function that concentrates the path integral's integration measure on those ${\cal M}_n$ for which the bulk metric is continuous at each $\Sigma_{M^i_-},\Sigma_{\tilde M^{\dagger i}_-}$. This will also result in various constraints relating the induced metrics on the splitting surfaces $\fixM_i, \tilde \fixM_i$, though the detailed form of such constraints depends on the particular pattern of cut-and-paste operations required to form a given ${\cal M}_n$.

\begin{figure}
\centering
\includegraphics[width=\linewidth, scale=0.2]{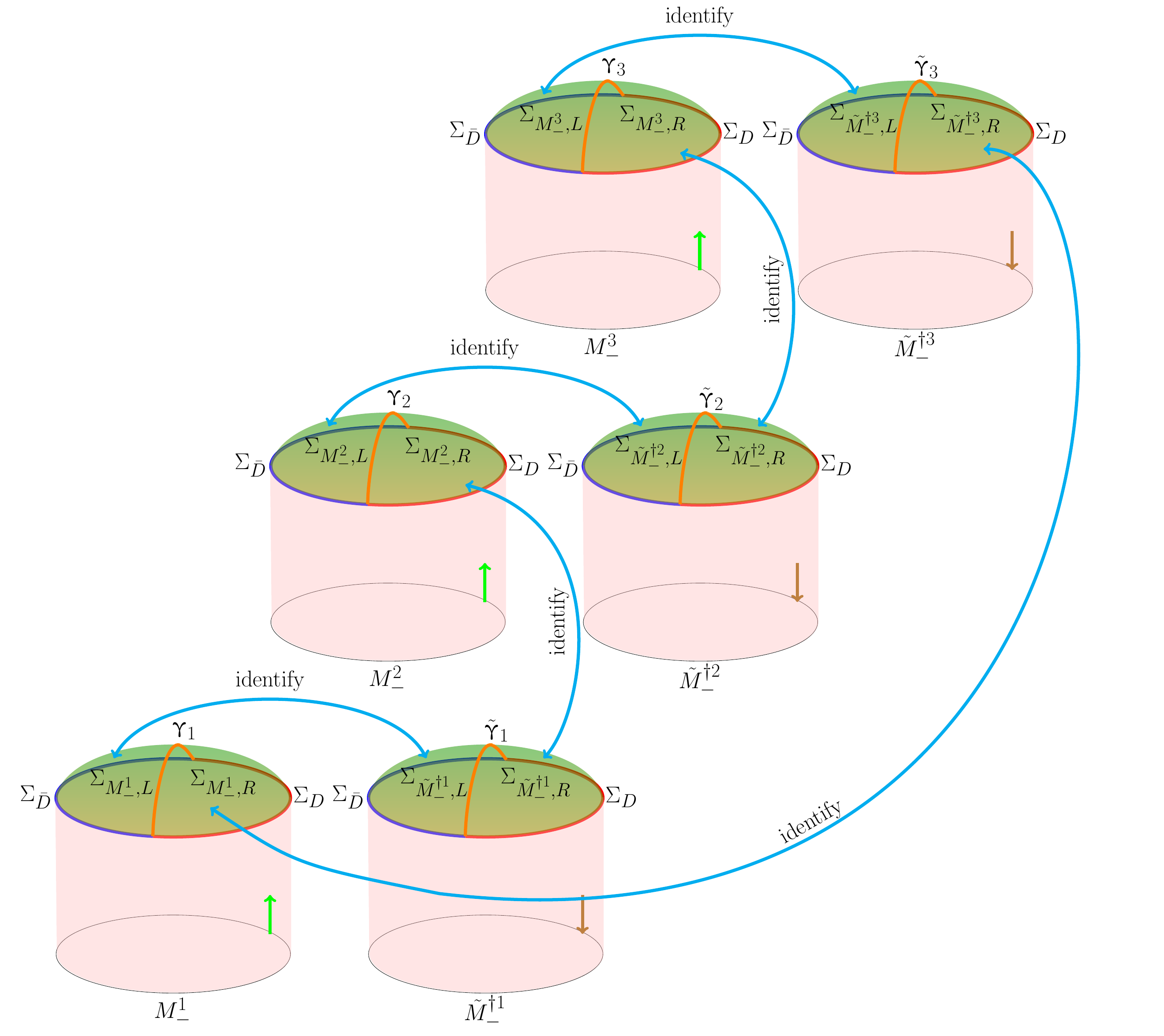}
\caption{A typical bulk configuration of the gravitational path integral that computes $\Tr [(\rho(D))^n]$ for $n=3$.}
\label{fig:Mn}
\end{figure}

It remains to further specify the path integral weight for ${\cal M}_n$.  Since the action is local, we can specify the weight for each region of ${\cal M}_n$ independently.  For regions away from $\Sigma_{M^i_-},\Sigma_{\tilde M^{\dagger i}_-}$, this is just the relevant weight $e^{\pm i S_L}$ for the corresponding region of $M^i_-,\tilde M^{\dagger i}_-$.  Indeed, we will use the same rule for regions that intersect $\Sigma_{M^i_-},\Sigma_{\tilde M^{\dagger i}_-}$ but which remain away from all splitting surfaces $\fixM_i, \tilde \fixM_i$.  In doing so, we should recall that the action $S_L$ controlling the weight for $M^i_-,\tilde M^{\dagger i}_-$ must include a Gibbons-Hawking term since such pieces each individually describe contributions to a path integral with fixed induced metric on $\Sigma_{M^i_-},\Sigma_{\tilde M^{\dagger i}_-}$. See figure \ref{fig:Mn} for an illustration of $\mathcal{M}_n$.

It now remains to specify the action for regions that include a splitting surface $\fixM_i, \tilde \fixM_i$.  For Einstein-Hilbert gravity we may follow the guiding principle that, while the above cut-and-paste operations will introduce singularities at the codimension-2 surfaces $\fixM_i, \tilde \fixM_i$, it is only the structure in the transverse two-planes to $\fixM_i, \tilde \fixM_i$ that will be singular.  As a result, it suffices to understand the result for two-dimensional ${\cal M}_n$ in detail, and then to simply integrate that result along the $\fixM_i, \tilde \fixM_i$.  Furthermore, in two dimensions one can use the Gauss-Bonnet theorem to write the integral of $\sqrt{-g} R$ over any region $\mathscr{U}$ in terms of the Euler character $\chi(\mathscr{U})$ and an integral of the extrinsic curvature over the  surface $\partial \mathscr{U}$ (which by definition avoids the codimension-2 singularities at  $\fixM_i, \tilde \fixM_i$).  While the Gauss-Bonnet theorem is most familiar in the Euclidean context, there is a generalization to complex two-dimensional spacetimes that in particular includes the case relevant to small regions
$\mathscr{U}$ around each $\fixM_i, \tilde \fixM_i$ where the metric is real and (at least at points away from $\mathscr{U}$) has Lorentz signature.   See \cite{Colin-Ellerin:2020mva} for details and \cite{Louko:1995jw}  for an earlier use of the complex Gauss-Bonnet theorem to evaluate the gravitational action at a related singularity; see also the examples in \cite{Colin-Ellerin:2021jev}.\footnote{\label{foot:HD} If one wishes to generalize this argument to include higher derivative corrections,  the complex Gauss-Bonnet theorem may no longer suffice to determine the action completely.  However, one may still follow the Lorentz-signature version of the minimal-subtraction prescription described in the appendices of \cite{Dong:2019piw}, together with the Legendre transforms also described in \cite{Dong:2019piw}.}

In summary, we may follow \cite{Colin-Ellerin:2020mva} in writing the desired path integral as
\begin{equation}\label{eq:skMn}
Z[{\cal B}_n] := \int_n [Dg] \, e^{i\, S},
\end{equation}	
where the subscript $n$ is a reminder that we integrate over spacetimes ${\cal M}_n$ of the form described above satisfying boundary conditions defined by ${\cal B}_n$.  In particular, the above action takes the form
\begin{equation}
\label{eq:TotalS}
S({\cal M}_n) = \sum_{i=1}^n \left[S_L(M^i_-) - S_L (\tilde M^{\dagger i}_-)\right]  + S_{\fixM},
\end{equation}
where $\fixM$ is the union of $\fixM_i, \tilde \fixM_i$ modulo the identifications induced by the manner in which the $M^i_-, \tilde M^{\dagger i}_-$ are pasted together to form ${\cal M}_n$.  For Einstein-Hilbert gravity, the contribution $S_{\fixM}$ is defined by using the complex Gauss-Bonnet theorem  and integrating along $\fixM$ as outlined above; see \cite{Colin-Ellerin:2020mva} for details.   As a reminder, for AdS Einstein-Hilbert gravity with AdS scale $\ell_{AdS}$ the Lorentzian action of each piece  is
\begin{equation}\label{eq:EHact}
\begin{split}
S_L[M_-] &=
	 \frac{1}{16\pi G_N} \,  \int_{M_-}\, d^{d+1} x\, \sqrt{-g} \left[ R + \frac{d(d-1)}{\ell_{AdS}^2} \right] + \frac{1}{8\pi G_N} \int_{{\cal B}_-}\, d^{d} x\, \sqrt{|\gamma|} \, K  \\
& \qquad \qquad
	+ \frac{1}{8\pi G_N} \int_{\Sigma_{M_-}}\, d^{d} x\, \sqrt{h} \, K + S_\text{ct}({\cal B}_-)\, ,
\end{split}
\end{equation}	
where ${\cal B}_-$ is the asymptotic boundary of the piece $M_-$ (including any Euclidean or complex-signature regions of this asymptotic boundary),  $S_\text{ct}$ denotes an appropriate set of counter-terms, and the full boundary of $M_-$ is $\partial M_- = {\cal B}_- \cup \Sigma_{M_-}$.

\subsection{Boundary Conditions at the Splitting Surface}
\label{sec:split}

In describing the bulk spacetimes ${\cal M}_n$ that contribute to our path integral, we have thus far glossed over one important detail.  The issue is that we should require the ${\cal M}_n$ to satisfy boundary conditions at $\fixM$ for which the above $S$ defines a good variational principle.  It was shown in \cite{Colin-Ellerin:2020mva} that this is indeed the case if, on each piece $M^i_-, \tilde M^{ \dagger i}_-$ and near any component of $\fixM$,  one introduces coordinates $y^I$ along $\fixM$ and $\tilde x^\pm$ in the transverse space (with the component of $\fixM$ lying at $\tilde x^\pm =0$)  and requires the metric in this region to take the form
\begin{equation}\label{eq:mxL}
\begin{split}
ds^2 &=
	\sigma(\tilde x^+,\tilde x^-)\, d\tilde x^+  d\tilde x^- + T \,\frac{(\tilde x^+ \,d\tilde x^--\tilde x^-\, d\tilde x^+)^2}{(\tilde x^+\tilde x^-)^{2-\hat m}} \\
&\qquad \qquad 	
	+ q_{IJ} \, dy^I dy^J + 2\, W_J \, dy^J \, \frac{\tilde x^+ d\tilde x^--\tilde x^- d\tilde x^+}{\left(\tilde x^+ \tilde x^-\right)^{1-\hat m}},\\
\text{with} & \\
\sigma&(\tilde x^+, \tilde x^-)
\equiv
	\hat{m}^2 (\tilde x^+ \tilde x^-)^{\hat{m}-1}\, ,\\
T&= \order{(\tilde x^+\tilde x^-)^{\frac{\alpha \hat{m}}{2}}} \,,\\
q_{IJ}
&=
	\order{(\tilde x^+\tilde x^-)^0 }\,,\\
W_J&= \order{(\tilde x^+\tilde x^-)^{\hat{m}}}\, ,
\end{split}
\end{equation}
for some $\alpha >1$ and some $\hat m >0$.  The metric coefficients $T, q_{IJ}, W_J$ depend smoothly on the $y^I$ coordinates.
Note that, following the conventions of \cite{Colin-Ellerin:2020mva}, $\sigma$ is positive for $\hat m=1$.  One should thus think of $\tilde x^\pm$ as analogues of the Minkowski-space coordinates $x \pm t$ as opposed to the more standard null coordinates $t\pm x$.

It remains to specify some further details.  For positive $\tilde x^\pm$ the fractional powers above are defined by taking the positive real root.  But for negative $\tilde x^+$ we take
$(\tilde x^+)^{\hat m} = e^{-{i \hat m \pi}} \,  \abs{\tilde x^+}^{\hat m}$
and for negative $\tilde x^-$ we define
$(\tilde x^-)^{\hat m} = e^{+{i\hat m \pi}} \,\abs{\tilde x^-}^{\hat m}$.

In addition, the notation $\order{(\tilde x^+\tilde x^-)^q}$ in principle allows terms of the form $(\tilde x^+\tilde x^-)^q f(\frac{\tilde x^+}{\tilde x^-})$ for any smooth $f$.  However, if there are $2k$ pieces $M^i_-, \tilde M^{\dagger i}_-$ that meet at the relevant component of $\fixM$ (so that this component is formed by identifying $2k$ splitting surfaces $\fixM_i, \tilde \fixM_i$), then we also require the metric functions $T,\,q_{IJ} ,\, W_J$  to involve only integer powers of $(\tilde x^\pm)^{\frac{1}{k}}$, except perhaps in the combination $\tilde x^+ \tilde x^-$.  In other words, we require these coefficients to be functions of the triple $((\tilde x^+)^{\frac{1}{k}},(\tilde x^-)^{\frac{1}{k}},\tilde x^+ \tilde x^-)$ such that these functions are analytic in the first two arguments in some neighborhood of the origin $\tilde x^+ = 0 = \tilde x^-$.  As remarked in \cite{Colin-Ellerin:2020mva}, such local analyticity is to be expected at any source-free regular point of the equations of motion and does not restrict the generality of any saddle points that we will find.  (Configurations invariant under the local ${\mathbb Z}_n$ replica symmetry will have $k=n$ at each component of $\fixM$ and will be further restricted by the condition that they involve only integer powers of $\tilde x^\pm$, again with the possible exception of their appearance in the combination $\tilde x^+ \tilde x^-$.)

\subsection{Real-time R\'enyi saddles}
\label{sec:saddles}
We now briefly review general features of the saddle points ${\cal M}_n$ of the gravitational R\'enyi path integral $Z[{\cal B}_n]$.  The fact that \cite{Colin-Ellerin:2020mva} showed the definitions of the previous section to lead to a good variational principle means that ${\cal M}_n$ is a stationary point when it satisfies the standard (and, in our conventions, Lorentz signature) Einstein equations away from $\Sigma_{M^i_-},  \Sigma_{\tilde M^{\dagger i}_-}$ together with two simple additional conditions on $\Sigma_{M^i_-},  \Sigma_{\tilde M^{\dagger i}_-}$.  The first condition is just that where $\Sigma_{M^i_-}$ is sewn to some $\Sigma_{\tilde M^{\dagger j}_-}$ (but away from $\fixM$), the codimension-1 extrinsic curvatures of these surfaces must agree (e.g., when both are computed using future-pointing normals).  The second condition is that for each component of $\fixM$ we must impose $k\hat m =1$, where $k, \hat m$ are the parameters defined in and below equation \eqref{eq:mxL}.  The latter condition comes from varying the action with respect to the area element on $\fixM$, and may be thought of as the condition that the Ricci scalar contain no delta-function at this surface.  As a result, that despite the singular form of \eqref{eq:mxL}, there remains some physical sense in which we might think of the metric as being in some sense``smooth.''

Recall that  $k \in \mathbb{Z}^+$, and also that the case $k=1$ describes a simpler construction for which splitting surfaces were not in fact required.  In this sense we have $k \ge 2$ at a nontrivial component of $\fixM$. So at a nontrivial such component, the saddle-point condition $k\hat m=1$ forbids $\hat m$ from being an integer.  Examination of \eqref{eq:mxL} then shows that the metrics for such saddles are always complex-valued.  Following \cite{Colin-Ellerin:2020mva}, we assume that the original real contour of integration can be deformed to pass through such complex saddles, though it would be useful to investigate this more carefully in the future.

We will focus below on saddles ${\cal M}_n$ that preserve both replica and conjugation symmetry. One consequence of replica symmetry is that all components of $\fixM$ have $k=n$.  Another is that, as noted below \eqref{eq:mxL}, there must be a region around $\fixM$ where the functions $T, q_{IJ}, W_I$ are analytic in $\tilde x^\pm$ up to functions of the product $\tilde x^+ \tilde x^-$.

Finally, when combined with conjugation symmetry, replica symmetry imposes a further constraint where any $\Sigma_{M^i_-}$ is sewn to some $\Sigma_{\tilde M^{\dagger j}_-}$.  At such loci, any quantity on $\Sigma_{M^i_-}$ must be the complex conjugate of the corresponding quantity on $\Sigma_{\tilde M^{\dagger j}_-}$.  In particular,  since our sewing conditions require the induced metrics on these surfaces to agree, it follows that the induced metrics on $\Sigma_{M^i_-},\Sigma_{\tilde M^{\dagger j}_-}$ must be real.  And since the above saddle-point conditions require the extrinsic curvatures to agree on the part of this seam away from $\fixM$, such extrinsic curvatures must be real as well.  We conclude that replica- and conjugation-invariant saddles have real initial data on each $\Sigma_{M^i_-}\setminus \fixM_i ,\Sigma_{\tilde M^{\dagger j}_-}\setminus \tilde \fixM_j$.  Since the relevant equations of motion are the Lorentz-signature Einstein equations, we expect that we can use this initial data to construct real Lorentz-signature spacetimes that we may call the domain of dependence of each
$\Sigma_{M^i_-}\setminus \fixM_i ,\Sigma_{\tilde M^{\dagger j}_-}\setminus \tilde \fixM_j$. This is clear when the induced metric on 
$\Sigma_{M^i_-}\setminus \fixM_i ,\Sigma_{\tilde M^{\dagger j}_-}\setminus \tilde \fixM_j$ is positive definite, in which case we have defined good Cauchy data on this surface.  Furthermore, in that case uniqueness of the initial value problem tells us
that this is in fact the desired saddle in the stated region.  In other words, replica- and conjugation-invariant saddles must be real in what we may call `the region spacelike separated from $\fixM.$' 

However, it remains to consider the case where the induced metric on $\Sigma_{M^i_-}\setminus \fixM_i ,\Sigma_{\tilde M^{\dagger j}_-}\setminus \tilde \fixM_j$ fails to be positive definite.  Such cases are not at all pathological, and in fact are easily generated from the positive-definite cases above.  In such cases, the fact that the saddle is real in the region spacelike separates from $\fixM$ means that a factor of $e^{iS}$ from some part of this region in a ket spacetime will exactly cancel against the corresponding $e^{-iS}$ factor from the corresponding region of a bra spacetime.  As a result,  without changing the action or the validity of the saddle we can deform the original 
$\Sigma_{M^i_-}\setminus \fixM_i ,\Sigma_{\tilde M^{\dagger j}_-}\setminus \tilde \fixM_j$ to an {\it arbitrary} surface spacelike separated from $\fixM$ that connects $\fixM$ with $\Sigma_D$.  In particular, the deformed surface may contain timelike or null regions, in which case the induced metric will fail to be positive definite. We will assume that all cases where the induced metric fails to be positive definite can be obtained in this way.  This is the one assumption foreshadowed in the introduction.  It was also implicit in \cite{Colin-Ellerin:2020mva}.

\subsection{Time independence of classical (annealed) swap entropies}
\label{sec:classical}

Having reviewed real-time gravitational R\'enyi path integrals and their saddles in section \ref{sec:review}, we are now ready to verify the claim that -- in any saddle that preserves replica and conjugation symmetry --  the splitting surface $\fixM$ can be treated as an extremal surface in a real Lorentz signature spacetime.  A critical point is the observation reviewed above that such saddles do in fact have real and Lorentz signature metrics in the domains of dependence of the surfaces
$\Sigma_{M^i_-}\setminus \fixM_i ,\Sigma_{\tilde M^{\dagger j}_-}\setminus \tilde \fixM_j$.

Indeed, we will now use this observation to introduce what we will call a real `shadow' of the replica wormhole described above.  To begin, recall that $\fixM$ partitions each boundary $\Sigma_{M^i_-}, \Sigma_{\tilde M^{\dagger j}_-}$ into a piece that ends on $\Sigma_D$ and a piece that ends on $\Sigma_{\bar D}$.  Let us refer to the pieces ending on $\Sigma_D$ as the `right' pieces
$\Sigma_{M^i_-, R}, \Sigma_{\tilde M^{\dagger j}_-, R}$ and the pieces
ending on $\Sigma_D$ as the `left' pieces
$\Sigma_{M^i_-, L}, \Sigma_{\tilde M^{\dagger j}_-, L}$.  Note that the initial data sets on any two right pieces are related by replica and conjugation symmetry, and that the same is true of the initial data sets on any two left pieces.

\begin{figure}
\centering
\includegraphics[scale=0.8]{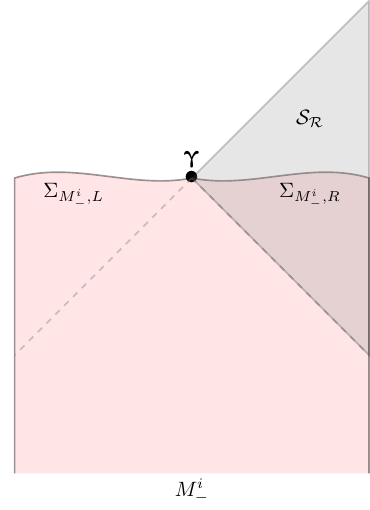}
\caption{The `right shadow' $\mathcal{S_R}$ (shaded in gray) is the maximal AdS-Cauchy development of $\Sigma_{M_-^i,R}$.  Note that $\mathcal{S_R}$ includes regions both to the past and to the future of $\Sigma_{M^i_-, R}$.  The region to the past also lies in $M_-^i$ (shaded pink), while the region to the future does not.   The intersection of $\mathcal{S_R}$ and $M_-^i$ has both shadings.  }
\label{fig:SR}
\end{figure}

Consider then the initial data on some $\Sigma_{M^i_-, R}$.  Since this data is real, the fact that the Lorentz-signature Einstein equations have a good initial value problem allows us to construct a unique solution ${\cal S_{R}}$ (the `right shadow') describing  the maximal AdS-Cauchy development\footnote{By an AdS-Cauchy development, we mean the analogue of a Cauchy development defined using the asymptotically AdS boundary conditions on $D$.  In other words, the surface $\Sigma_{M^i_-, R}$ is allowed to be an AdS-Cauchy surface for this development in the sense of \cite{Wall:2012uf}.} of $\Sigma_{M^i_-, R}$.  Furthermore, ${\cal S_{R}}$ must agree with the saddle point solution ${\cal M}_n$ on the domain of dependence of $\Sigma_{M^i_-, R}$ in $M^i_-$.   In particular,  near $\fixM$ the shadow ${\cal S_{R}}$ can be described by \eqref{eq:mxL}, where to be definite we choose $\tilde x^\pm >0$ in ${\cal S_{R}}$ (corresponding to our choice to think of this as the right wedge). However, since it lies in $M^i_-$ the latter region exists only to the past of $\Sigma_{M^i_-, R}$, while ${\cal S_{R}}$ includes a region to the future as well; see figure \ref{fig:SR}.

Now, in the region $\tilde x^\pm >0$ we are free to introduce new coordinates $\tilde X^\pm = (\tilde x^\pm)^{1/n}$.  Recalling that $\hat m = 1/n$, in terms of such coordinates the metric \eqref{eq:mxL} becomes just
\begin{equation}\label{eq:mXL}
\begin{split}
ds^2 &=
	d\tilde X^+  d\tilde X^- + n^2 T \,\frac{\left(\tilde X^+ \,d\tilde X^--\tilde X^-\, d\tilde X^+\right)^2}{\tilde X^+ \tilde X^-} \\
&\qquad \qquad 	
	+ q_{IJ} \, dy^I dy^J + 2n\, \tilde W_J \, dy^J \, \left(\tilde X^+ d\tilde X^--\tilde X^- d\tilde X^+\right),\\
\text{with} & \\
T&= \order{(\tilde X^+\tilde X^-)^{\frac{\alpha }{2}}} \,,\\
q_{IJ}
&=
	\order{(\tilde X^+\tilde X^-)^0 }\,,\\
\tilde W_J&= \frac{W_J}{\tilde X^+\tilde X^-} =  \order{(\tilde X^+\tilde X^-)^0}\, .
\end{split}
\end{equation}
Furthermore, the conditions on the coefficients $T, q_{IJ}, W_I$ now state that they should be functions of the triple $\left( (\tilde X^+)^n, (\tilde X^-)^n, \tilde X^+\tilde X^-\right)$ that are analytic in the first two arguments, at least in some neighborhood of $\tilde X^+ = \tilde X^-=0$.  As a result, the metric admits an extension $\hat {\cal S}_R$ to at least some negative values of both $\tilde X^+$ and $\tilde X^-$.

We would like to control the form of the extension  $\hat {\cal S}_R$ .  The idea is to use the fact that we are interested in the saddle ${\cal M}_n$ which solves the Einstein equations for $\tilde X^\pm \neq 0$.

In particular, in the Euclidean context ref. \cite{Dong:2019piw} found a power series solution near the splitting surface which (at least at the level of function counting) had sufficient freedom to accommodate general smooth boundary conditions and which obeyed boundary conditions given by the Wick rotation of \eqref{eq:mxL}.  Since the formal manipulation of power series is unchanged by Wick rotations, this immediately provides a similarly general solution  to the Einstein equations in our context.  We will thus now assume that any actual saddle ${\cal M}_n$ defined by sufficiently smooth boundary conditions can be described by such a power series.  While it would be interesting to prove this statement rigorously, that task is beyond the scope of this work.  We therefore leave it for appropriate mathematicians to investigate in the future.

In terms of $\tilde X^\pm$, and after the relevant Wick rotation and imposing replica symmetry and the condition $\hat m=1/n$, the power series solution from \cite{Dong:2019piw} may be written as sums over integers $p,q,s$ of the form
\begin{equation}
\label{eq:T}
T\sim \sum_{p, q, s=0 \atop p q>0, \text { or } s>0}^{\infty} T_{p q s} (\tilde{X}^{+})^{pn} (\tilde{X}^{-})^{qn} (\tilde{X}^+ \tilde{X}^-)^{s},
\end{equation}
\begin{equation}
\label{eq:W}
\tilde W_{I} \sim   \sum_{p, q, s=0}^{\infty} W_{I, p q s} (\tilde{X}^{+})^{p n} (\tilde{X}^{-})^{q n} (\tilde{X}^+ \tilde{X}^-)^{s},
\end{equation}
\begin{equation}
\label{eq:q}
q_{IJ}\sim \sum_{p, q, s=0}^{\infty} q_{IJ, p q s} (\tilde{X}^{+})^{pn} (\tilde{X}^{-})^{qn} (\tilde{X}^+ \tilde{X}^-)^{s}.
\end{equation}

In particular, the restriction $ p q>0, \text { or } s>0$ in the sum for $T$ means that for $n > 1$ the leading term will be $\tilde X^+ \tilde X^-$.  The coefficient $\frac{T}{\tilde X^+ \tilde X^-}$ in \eqref{eq:mXL} is thus smooth at $\tilde X^\pm=0$.

Using the above power series expansion, we may choose the extension  $\hat {\cal S}_R$  to negative $\tilde X^\pm$ to be arbitrarily smooth. And since it satisfies the Einstein equations on ${\cal S}_R$, we may choose the extension to satisfy the Einstein equations as well.\footnote{For this it suffices to first smoothly extend the initial data on some Cauchy surface and to then solve the Einstein equations.}  We also take $\hat {\cal S}_R$ to satisfy standard causality assumptions such as those used in \cite{Wall:2012uf,Headrick:2014cta}, perhaps at the cost of limiting the extension to negative values of $\tilde X^\pm$ that are very close to $\fixM$.  In particular, there is no need for the extension  $\hat {\cal S}_R$ to be maximal in any sense.

By counting powers of $\tilde X^\pm$ is also easy to see that the codimension-2 extrinsic curvature of $\fixM$ in $\hat {\cal S}_R$  must vanish in $\hat {\cal S}_R$.\footnote{Aside from the terms independent of $\tilde X^\pm$, for integer $n>1$  almost every term in \eqref{eq:mXL} is at least quadratic in $\tilde X^+ \tilde X^-$.  The single exception is the leading contribution from $W_J$.  But the term associated with the leading contribution from $W_J$ is invariant under  $\tilde X^\pm \rightarrow -\tilde X^\pm$ and so cannot contribute to the extrinsic curvature of the set $\fixM$ at $\tilde X^\pm =0$. }  Indeed, for even $n$ comparing \eqref{eq:mXL} with
\eqref{eq:T}, \eqref{eq:W}, and \eqref{eq:q} shows that every term in \eqref{eq:mXL} is even under $\tilde X^\pm \rightarrow -\tilde X^\pm$, so we are free to take $\hat {\cal S}_R$  to have an exact ${\mathbb Z}_2$ symmetry about $\fixM$ when $n$ is even.\footnote{\label{foot:shadow} For odd $n$, the fact that $\tilde X^\pm = (\tilde x^\pm)^{1/n}$ maps positive real $\tilde x^\pm$ to positive real $\tilde X^\pm$ and also maps negative real $\tilde x^\pm$ to negative real $\tilde X^\pm$ means that in that case we can instead take the extension $\hat {\cal S}_R$ in the region where $\tilde X^\pm$ are both negative to coincide with the `left shadow' ${\cal S}_L$ defined in analogy with ${\cal S}_R$ but using the maximal Cauchy development of $\Sigma_{M^i_-, L}$. Both properties are easy to see in the case where ${\cal M}_n$ is the Wick rotation of some ${\cal M}_n^E$, as comparison with \cite{Colin-Ellerin:2020mva} shows that $\hat{\cal S}_R$ can be defined by a different Wick-rotation of ${\cal M}_n^E$ that preserves smoothness.  In effect, this is the Wick-rotation that would be obtained by interpreting the smooth ${\cal M}_n^E$ as an $n=1$ geometry that computes the norm of some pure state by slicing it into two pieces, representing a single bra and a single ket, each of which contains $n/2$ replicas and such that the two pieces are related by a ${\mathbb Z}_2$ symmetry that complex-conjugates all sources.  See figures in appendix \ref{app:figs}.}

For integer $n>1$ we have now argued that $\fixM$ has vanishing extrinsic curvature in an arbitrarily smooth spacetime $\hat {\cal S}_R$ which satisfies standard causality assumptions as well as the vacuum Einstein equations (and thus also the null convergence condition).  In other words, the associated matter stress tensor vanishes identically and thus satisfied the null energy condition.  Note that this result at $\tilde X^\pm=0$ follows from smoothness and from the corresponding result at positive $\tilde X^\pm$.  The results of \cite{Wall:2012uf,Headrick:2014cta} then require $D$ to be causally inaccessible from $\fixM$.

This is essentially the desired result, though we should carefully state what this means for the original saddle ${\cal M}_n$ on which the metric in some regions is complex-valued.  In order to do so, let us first return to $\hat {\cal S}_n$ and note that the boundary of the future of $\fixM$ cannot intersect $D$, and neither can the boundary of the past.  This is because $D$ is an open set on the asymptotically AdS boundary, so such an intersection would require that $D$ also intersect the interior of the causal past or future, contradicting the statement that it is causally inaccessible from $\fixM$.

On the other hand, since the Cauchy surface $\Sigma_{M^i_-, R}$ intersects $D$, and since standard asymptotically AdS boundary conditions hold at $D$, the maximal AdS-Cauchy development ${\cal S_{R}}$ of $\Sigma_{M^i_-, R}$ must contain part of $D$.  In fact, since $\Sigma_{M^i_-, R}$  intersects $D$ on a Cauchy surface for $D$, it intersects every connected component of $D$.  Thus ${\cal S_{R}}$  in fact contains at least part of every such connected component.

However,  the boundary of the future of $\fixM$ in $\hat {\cal S}_R$ is precisely
the boundary of ${\cal S}_R$.  Since this boundary cannot intersect $D$, it follows that any connected component of $D$ which intersects ${\cal S}_R$ is in fact fully contained in ${\cal S}_R$.  And since this was the case for all such components, we in fact conclude that $D$ is fully contained in the maximal Cauchy development of $\Sigma_{M^i_-, R}$.

This, then,  is the desired result.  For brevity, we will  refer to it as stating that `$D$ is causally inaccessible from $\fixM$ in ${\cal M}_n$'.  This phrasing is motivated by the fact that, in any real Lorentz-signature globally-hyperbolic spacetime, any region $R$ of a Cauchy surface $\Sigma$ must be causally inaccessible from the maximal Cauchy development of the complementary region $\bar R = \Sigma \setminus R$.

\begin{figure}
\centering
\includegraphics[scale=0.8]{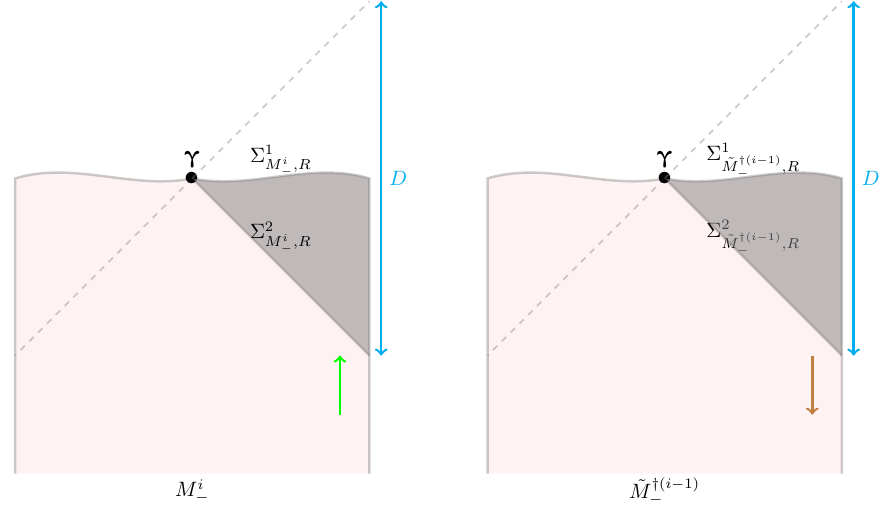}
\caption{
In every saddle, the sheets $M_-^i$ and $\tilde M_-^{\dagger (i-1)}$ are sewn together by identifying an achronal surface $\Sigma_{M_-^i,R}$ in $M_-^i$ with an achronal surface $\Sigma_{\tilde M_-^{\dagger (i-1)},R}$ in  $M_-^{\dagger (i-1)}$.  These surfaces have boundaries consisting of $\fixM$ and $\Sigma_D$; i.e., $\partial \Sigma_{M_-^i,R} = \partial \Sigma_{\tilde M_-^{\dagger (i-1)},R} = \fixM \cup \Sigma_D$.  However, the action of saddles with replica and conjugation symmetry is independent of the choice of this achronal surface, and the action is also independent of the choice of Cauchy surface $\Sigma_D$ for $D$.  As a result, without changing the action we may deform any such $\Sigma^1_{M_-^i,R}$ and $\Sigma^1_{\tilde M_-^{\dagger (i-1)},R}$ to surfaces $\Sigma^2_{M_-^i,R}$ and $\Sigma^2_{\tilde M_-^{\dagger (i-1)},R}$  on the past lightcone of $\fixM$.   This makes manifest that the contribution of our saddle is independent of any choices within $D$. }
\label{fig:lightcone}
\end{figure}

However, the key point is that the solution is real on the maximal AdS Cauchy development of $\Sigma_{M^i_-, R}$, $\Sigma_{\tilde M^{\dagger j}_-, R}$.  But our symmetries require the solution on any AdS-Cauchy development of $\Sigma_{M^i_-, R}$ to be the complex conjugate of that on the corresponding development of  $\Sigma_{\tilde M^{\dagger j}_-, R}$.  Thus the solutions on these developments agree and their contributions cancel in \eqref{eq:TotalS}.  Thus we obtain the same action if we choose $\Sigma_{M^i_-, R}$, $\Sigma_{\tilde M^{\dagger j}_-, R}$ to lie on the past light cone of $\fixM$.  But this has the effect of removing $D$ entirely from the boundary ${\cal B}_n$ and thus makes manifest that the contribution of our saddle is independent of any choices within $D$.  See figure \ref{fig:lightcone}.

\subsection{Generalizations to include matter}

The above arguments establish our result for vacuum Einstein-Hilbert gravity.  Since we considered spacetimes of arbitrary dimension, we may then use Kaluza-Klein to deduce corresponding results for  of Einstein-Hilbert gravity coupled to appropriate Kaluza-Klein matter, and in particular for the truncation of such theories to modes that preserve any symmetries of the internal manifold.  Since non-linearities contribute only higher-order terms to our expansions above, this suggests that the exact form of the matter couplings is not relevant.  We therefore expect our result to hold for arbitrary two-derivative theories of matter minimally-coupled to Einstein-Hilbert gravity for which the matter sector satisfies the null energy condition.  However, we leave a detailed proof of this generalization for future work.


\section{Time independence of quantum-corrected (annealed) swap R\'enyis}
\label{sec:quantum}

Having established time-independence of gravitational annealed swap R\'enyis at leading order in $G_N$, it is of interest to consider quantum corrections.  We focus here on corrections associated with matter fields, leaving aside subtleties associated with quantum corrections from gravitons.  We expect that such subtleties can be dealt with using the techniques of \cite{Dong:2017xht}, but we will not attempt to do so here.

In the classical case, we found that we could use a smooth real Lorentz signature `shadow spacetime' to describe the area of any surface in ${\cal M}_n$ that was causally inaccessible from $\fixM$.  Moreover, replica symmetry of the classical replica saddle ${\cal M}_n$ led to a power series expansion of this area near $\fixM$ which showed $\fixM$ to be extremal.  The null curvature condition on ${\cal M}_n$ then required $D$ to be contained in the maximal AdS-Cauchy development of $\Sigma_{M_-^i}$, a situation that we summarize by saying that $D$ is  causally inaccessible from $\fixM$.

We would like to simply extend this argument to include quantum corrections, replacing the null curvature condition with the quantum focussing condition (QFC) of \cite{Bousso:2015mna}.  This condition involves the entropy $S_{QFT}$ of quantum fields on ${\cal M}_n$, which may be subtle in regions where the metric is complex.  But it should have familiar properties in the regions where ${\cal M}_n$ is real and Lorentz signature.

Let us begin by studying the first-order quantum corrections about a classical saddle replica ${\cal M}_n$.  Then since the smooth shadow ${\cal S}_R$ is related by Cauchy evolution to a subset of a single sheet of ${\cal M}_n$.,
we may again describe the desired matter entropy $S_{QFT}$ as a function on ${\cal S}_R$.

The remaining steps in the classical argument then followed from properties of power series expansions near $\fixM$.  These expansions were motivated by requiring the relevant quantities to satisfy the classical equations of motion.  But to apply analogous reasoning in the quantum case we would need to understand the power series expansion for $S_{QFT}$.  Since we do not have classical equations of motion to solve for $S_{QFT}$, we will need to pursue another approach.

We describe two such approaches below in section \ref{sec:Wick} and \ref{sec:holographic}.  Each alone addresses only special cases. The first explicitly assumes that ${\cal M}_n$ can be Wick-rotated to define a smooth Euclidean (or perhaps complex) manifold, while the second works directly in Lorentz signature but assumes the QFT to be holographic and works in the approximation where $S_{QFT}$ can be computed using the classical Hubeny-Rangamani-Takayanagi prescription in the associated higher-dimensional spacetime.  However, after presenting these arguments, we will show in section \ref{sec:combine} that they can be combined to argue for general Lorentz-signature matter QFTs that the entropy
$S_{QFT}$ is in fact extremal on $\fixM$, and thus that $D$ is causally inaccessible from $\fixM$ as desired.  We then comment briefly on how the argument extends to higher order quantum corrections.

\subsection{Argument via Wick rotation}
\label{sec:Wick}

As stated above, for our first approach  we explicitly assume that ${\cal M}_n$ can be analytically continued to define a smooth Euclidean (or perhaps complex) manifold ${\cal M}^E_n$ by Wick rotating $\tilde x^\pm$ to complex coordinates $v, \bar v$ and defining the smooth coordinate $z=v^n, \bar z  = \bar v^n$, with the understanding that we rotate the timelike coordinate $\tilde  x^+ - \tilde x^-$ into the lower half plane in the ket spacetime and that we rotate $\tilde x^+ - \tilde x^-$ into the upper half plane in the bra spacetime.  Indeed, the coordinates $z, \bar z$ are just the analogous Wick rotation of $\tilde X^\pm$.  As a result, the metric \eqref{eq:mXL} and the expansions \eqref{eq:T}, \eqref{eq:W}, \eqref{eq:q} show that -- if the spacetime is sufficiently analytic for the Wick rotation to be defined -- the result will be smooth at the origin $z, \bar z =0$.

Furthermore, since Wick rotation preserves the replica symmetry that was assumed for the real-time saddle ${\cal M}_n$, the resulting ${\cal M}^E_n$  is replica-invariant.  This is clearly compatible with the Wick-rotation of the expansions \eqref{eq:T}, \eqref{eq:W}, \eqref{eq:q}, which are manifestly invariant under $z \rightarrow e^{\frac{2\pi i}{n}}z$ after the above rotation of $\tilde X^\pm$ to $z, \bar z$.

\begin{figure}
\centering
\includegraphics[width=\linewidth]{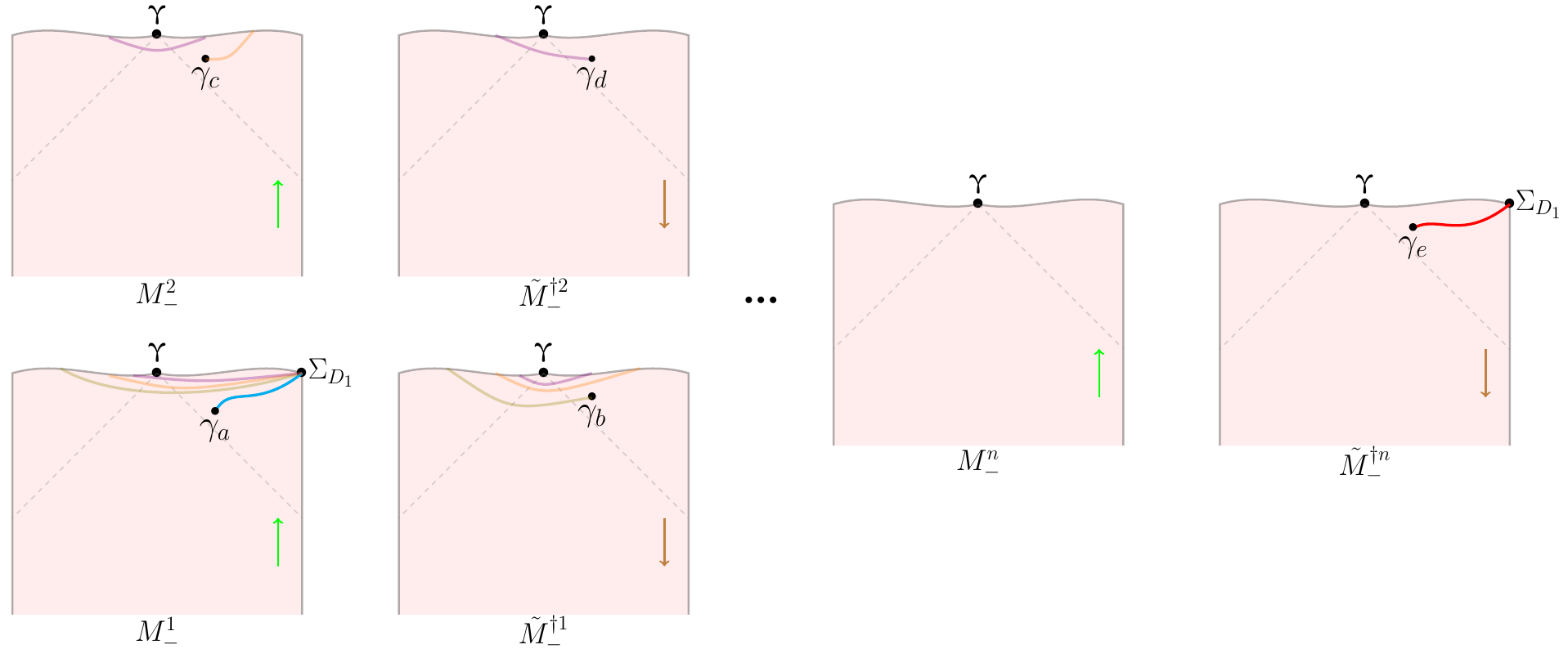}
\caption{After choosing a particular replica $D_1$ of $D$ and an associated Cauchy surface $\Sigma_{D_1}$, one can find an achronal surface $\Sigma_\gamma$ from some codimension-2 surface $\gamma$ to $\Sigma_{D_1}$ only when $\gamma$ is in the domain of dependence of $\Sigma_{M^1_-,R}$ or $\Sigma_{\tilde M^{\dagger n}_-,R}$ , but not in the domain of dependence of $\Sigma_{M^i_-,R}$ for $i \neq 1$ or $\Sigma_{\tilde M^{\dagger j}_-,R}$ for $j \neq n$. The blue and red surfaces are such $\Sigma_\gamma$ in the successful cases.   We also show typical surfaces (orange, brown, and magenta) that would connect other regions to $\Sigma_D$.  These latter surfaces must pass through regions with complex metric where achronality is not defined. Here the $M^i_-, \tilde M_-^{\dagger j}$ are sewn together as in figure \ref{fig:Mn}.  Thus, for example, the orange surface can be traced from $\gamma_c \subset M_-^2$ across the seam at the upper right of this sheet into $M_-^{\dagger 1}$, across that sheet moving to the left through the complex region to the past of $\fixM$, across the seam at the upper left of $M_-^{\dagger 1}$ into $M_-^1$, and finally across $M_-^1$ moving to the right through the complex region to the past of $\fixM$ to reach $\Sigma_D$. }
\label{fig:Sbulk}
\end{figure}

We may thus proceed to discuss the analytic continuation of $S_{QFT}$ to ${\cal M}^E_n$.  Let us begin by further discussing $S_{QFT}$ in the real-time saddle ${\cal M}_n$.  Recall that $S_{QFT}[\gamma]$ is the entropy on an achronal surface $\Sigma_\gamma$ stretching from $\gamma$ to a Cauchy surface $\Sigma_D$ for some replica of $D$ on the boundary. Furthermore, the notion of achronal surface is nominally defined only in region where the metric is real and Lorentz signature.  We will call this replica $D_1$ below and take it to lie on the boundary of $M^1_-$.  As a result, with our conventions for sewing together bra and ket spacetimes to make ${\cal M}_n$, this $D_1$ also lies on the boundary of $\tilde M_-^{\dagger n}$ (and not on the boundary of $\tilde M_-^{\dagger 1}$). Given this choice, one can generally find achronal surfaces $\Sigma_\gamma$ of the above form  when $\gamma$ in the domain of dependence of $\Sigma_{M^1_-,R}$ or $\Sigma_{\tilde M^{\dagger n}_-,R}$, but not when it lies in the domain of dependence of $\Sigma_{M^i_-,R}$ for $i \neq 1$ or $\Sigma_{\tilde M^{\dagger j}_-,R}$ for $j \neq n$; see figure \ref{fig:Sbulk}.

On the other hand, analytic continuation is trivial when $z$ is real and positive, as $z=\bar z >0$  is the surface $\Sigma_{M^1_-,R}$ on which the time coordinate $\tilde  x^+ - \tilde x^-$ vanishes.  Thus the surface lies in both ${\cal M}_n$ and ${\cal M}^E_n$.

Furthermore, on $\Sigma_{M^1_-,R}$ we may compute $S_{QFT}[\gamma]$ by using a second replica trick.  By this we mean that we take $m$ copies of the Euclidean spacetime ${\cal M}^E_n$ on which we wish to define the matter entropy $S_{QFT}$, cut them open, and then sew them back together using replica boundary conditions.  This introduces a conical singularity at $\gamma$, but since we are only using this second replica trick to compute the matter entropy of a QFT on a fixed background we do not otherwise alter the metric (i.e., there is no `backreaction' from this new conical singularity).  Computing the matter QFT path integral over the $m$-replica spacetimes and taking an appropriate $m\rightarrow 1$ limit gives $S_{QFT}$.  Furthermore  this second replica trick calculation can be performed for any $\gamma$ in ${\cal M}^E_n$ satisfying the homology constraint and so can be used to extend the definition of $S_{QFT}$ to all of ${\cal M}^E_n$.  Indeed, since there is no obstacle to the replica-trick computation giving an analytic result, we take it to be the desired analytic extension.

Note that this new replica trick explicitly preserves the ${\mathbb Z}_n$ replica-symmetry of the original ${\cal M}^E_n$.  We have thus defined $S_{QFT}$ as an analytic replica-symmetric function on ${\cal M}^E_n$.  As a result, given
any codimension-2 surface $\gamma$ about which this ${\mathbb Z}_n$ acts as a rotation,
first-order variations of $S_{QFT}[\gamma]$ must be invariant under such rotations and must thus vanish.

Applying this argument to the replica-invariant splitting surface $\fixM$ shows that $S_{QFT}$ is extremal on $\fixM$ as desired.  Returning to Lorentz signature and assuming the quantum focussing condition\footnote{Since we consider only first-order backreaction to Einstein-Hilbert gravity, one might think that one need only assume the quantum null energy condition (QNEC) of \cite{Bousso:2015mna}.  This is true when ${\cal M}_n$ spacetime dimension 3 or less.  But in higher dimensions the QNEC is not generally well-defined in curved space due to curvature-dependent UV divergences that depend on the choice of renormalization scheme \cite{Fu:2017evt, Akers:2017ttv}.} \cite{Bousso:2015mna} on the shadow ${\cal S}_R$, it immediately follows that this closed inequality holds on the closure of ${\cal S}_R$, and thus on the relevant outgoing null congruences from $\fixM$.  As a result, the generalized second law must hold on this null congruence \cite{Bousso:2015mna}.  We may then use the argument of \cite{Engelhardt:2018kcs} that, as in the classical case, $\fixM$ must be causally separated from $D$.

Unitarity of the bulk QFT then allows us to again deform each $\Sigma_{M^i_-,R}$ to the past light cone of $\fixM$ without changing $S_n(D)$.  As before, this makes manifest that $S_n(D)$ is independent of the choice of either $\Sigma_D$ or any sources within $D$.

\subsection{Argument for holographic matter}
\label{sec:holographic}

We argued in section \ref{sec:Wick} above that, when ${\cal M}_n$ can be appropriately Wick-rotated to a Euclidean spacetime ${\cal M}^E_n$, our matter entropy $S_{QFT}$ is stationary to first-order variations about $\fixM$.  We now give a different argument that works directly in the original real-time saddle ${\cal M}_n$ when the matter QFT happens to be holographic and $S_{QFT}$ is computed using the classical Hubeny-Rangamani-Takayanagi (HRT) prescription in the associated higher-dimensional asymptotically-AdS spacetime ${\cal N}_n$ having boundary ${\cal M}_n$.  For simplicity we also assume that it is sufficient to treat ${\cal N}_n$ as a solution to the vacuum Einstein equations (with a cosmological constant).  See appendix \ref{app:RSHRT} for a simple example that illustrates the discussion below.

Let us discuss relevant features of ${\cal N}_n$ and the associated HRT prescription.  Recall that ${\cal M}_n$ had an asymptotically-AdS boundary.  But if ${\cal M}_n$  were the boundary of a smooth bulk spacetime ${\cal N}_n$, then of course $\partial {\cal M}_n$ would be empty.  The resolution was explained in \cite{Aharony:2010ay}, which is to understand that the boundary of ${\cal M}_n$ must extend into the bulk as a dynamical object on which the bulk spacetimes can in some sense be said to end as well.  For example, in simple cases this internal `boundary' may be a string-theoretic orbifold or orientifold.  More generally, at a phenomenological level one can simply model the object (whatever it may be) as an  `end-of-the-world brane' \cite{Takayanagi:2011zk}.

The key point for our purposes is the implication for the form of the homology constraint that should be satisfied by HRT surfaces $\gamma_{\cal N}$ for $S_{QFT}[\gamma]$.  Recall that this is the entropy in ${\cal M}_n$  on a partial Cauchy surface $\Sigma_\gamma$ stretching from $\gamma$ to $\Sigma_D$.  Although one may still say that $\gamma_{\cal N}$ must be homologous to $\Sigma_\gamma$, if one thinks of the bulk as containing an end-of-the-world brane this is now homology in the sense of manifolds with boundary.  In other words, one requires only that there be a bulk surface $\Sigma_{\cal N}$ for which $\partial \Sigma_{\cal N}$ is $\gamma_{\cal N} \cup \Sigma_\gamma$ up to additional contributions that coincide with the end-of-the-world brane.  In certain microscopic descriptions this is clear from the fact that the end-of-the-world brane is really just a place where the bulk spacetime pinches off smoothly, but dualities require it to be true more generally.  Perhaps the more fundamental point is that the end-of-the-world brane is dynamical, so that points on its world-volume behave in much the same way as other points in the bulk.

Let us now discuss the form of ${\cal N}_n$ in more detail.  Note that the relationship of ${\cal N}_n$  to ${\cal M}_n$ is directly analogous to that of ${\cal M}_n$ to ${\cal B}_n$.  It is true that the metric on ${\cal M}_n$  is complex in regions that we might say lie to the past of $\fixM$ while the metric on ${\cal B}_n$ is real, but this point will not affect the discussion.  In particular, we have taken ${\cal M}_n$ to preserve all of the symmetries of ${\cal B}_n$, so we will assume ${\cal N}_n$ to preserve these symmetries as well.  The same argument as in section \ref{sec:saddles} then requires the metric on  ${\cal N}_n$  to be real and Lorentz signature in regions causally inaccessible from $\fixM_{\cal N}$, and the asymptotic form of the metric near $\fixM_{\cal N}$ will be again follow \eqref{eq:mxL}, \eqref{eq:mXL}.
We may thus construct a shadow ${\cal S}^{\cal N}_R$ and an extended shadow $\hat {\cal S}^{\cal N}_R$ of ${\cal N}_n$ using the recipe for ${\cal S}_R$ and $\hat {\cal S}_R$ from section \ref{sec:classical}.  We may also note that $\fixM_{\cal N}$ is homologous to $\Sigma_\gamma$ for $\gamma =\fixM$ in the sense defined above (since the relation between these surfaces is analogous to the relation between $\fixM$ and $\Sigma_D$).

As a result, by the same logic as was to discuss $\fixM$ in that section \ref{sec:classical}, we find that the replica-invariant surface $\fixM_{\cal N}$ in ${\cal N}_n$ must be extremal in ${\cal S}^{\cal N}_R$.   Using this argument in the bulk of ${\cal N}_n$  tells us that $\fixM_{\cal N}$ is extremal under bulk variations.  So since it is homologous to $\Sigma_\gamma$ for $\gamma = \fixM$, it is a candidate HRT  surface for $\Sigma_\fixM$ (i.e., for the surface $\Sigma_\gamma$ with $\gamma=\fixM$).  Let us thus follow section \ref{sec:classical} in assuming it to be the smallest such surface, in which case it must be the actual HRT surface for $\Sigma_\fixM$.

However, since the extremality argument used only properties of ${\cal N}_n$ that are also true of ${\cal M}_n$ -- an in particular properties of ${\cal M}_n$ that were already used in section \ref{sec:classical}  -- we may also conclude $\fixM_{\cal N}$ to be extremal with respect to variations of its boundary $\fixM \subset {\cal S}_R$.  Since we work in the approximation where $S_{QFT}[\gamma]$ is proportional to the area of the associated HRT surface in ${\cal N}_n$, it follows that  $S_{QFT}[\gamma]$  is stationary to first order about the surface $\gamma = \fixM$.  But this $\fixM$ was also a classical extremal surface, so this tells us that it is in fact a quantum extremal surface. If we assume the quantum focussing condition we may thus again repeat precisely the arguments at the end of section \ref{sec:Wick} to conclude that $\fixM$ is causally inaccessible from $D$.

\subsection{General Argument}
\label{sec:combine}

The holographic argument in section \ref{sec:holographic} served mainly to illustrate the general point made at the very beginning of our discussion of quantum corrections.  In particular, once some control was obtained over the asymptotic form of $S_{QFT}[\gamma]$ near the surface $\gamma = \fixM$, we were able to show stationarity of $S_{QFT}$ at $\fixM$ using precisely the same argument given for stationarity of the area at $\fixM$ in section \ref{sec:classical}.  The assumption that the QFT was holographic served only to allow us to extract the desired asymptotic expansion from the existing literature.

Furthermore, in both sections \ref{sec:classical} and \ref{sec:holographic}, the desired asymptotic expansion was obtained by finding a self-consistent power series solution to the appropriate equations of motion.  And in fact this was done by noting that Wick rotation is straightforward to any order in a series expansion, so that it was sufficient to transcribe the series solutions described for the Euclidean context in \cite{Dong:2019piw}.  Despite the use of this Wick rotation, the resulting series should provide a good asymptotic expansion for the desired quantity in any Lorentz-signature theory, even in the presence of non-analytic sources.

We thus wish to now implement the analogous steps in a general quantum field theory.  In that context the (variational) derivatives of $S_{QFT}$ may again be said to be described by `equations of motion' which related them to certain correlation functions defined by the appropriate modular Hamiltonian and the stress tensor; see e.g. \cite{Rosenhaus:2014zza}.  In principle, it should be possible to use this structure together with stress tensor conservation to construct the desired power series description of variations of $S_{QFT}$ about $\fixM$.   In practice, of course, this is a highly non-trivial task.

Luckily, as described above, the general form of this expansion must be the same whether or not the theory admits a Wick rotation to Euclidean signature.  Let us therefore assume that it does, and the rotation can be taken to have the form described in \ref{sec:Wick} above.  Then the expansion can be Wick rotated as well, and as in section \ref{sec:Wick} it must be the series expansion of some analytic replica-invariant functional $S_{QFT}$ on an analytic Euclidean (or complex) spacetime.  In particular, in this case both $S_{QFT}$ and the spacetime are analytic functions of the $z, \bar z$ obtained from Wick rotation of $\tilde X^\pm$.  Analyticity and replica symmetry then require $S_{QFT}$ to be stationary under first variations of $z, \bar z$ about $\fixM$, and thus also require
$S_{QFT}$ to be stationary under first variations of $\tilde X^\pm$ about $\fixM$.  But since such variations are determined by the series expansion, and since we argued this expansion to have the same form in the general case as in the analytic case, $S_{QFT}$ will be stationary on $\fixM$ for any QFT.  By assuming the quantum focussing condition, we may then again argue as above that $\fixM$ will remain causally inaccessible from $D$ under back-reaction from first-order quantum corrections.



%

\subsection{Higher order quantum corrections}
\label{sec:HOQC}

The argument given above clearly extends to higher order quantum corrections.  Working out the detailed expansions will become more cumbersome at higher orders, as one must take into account higher and higher levels of back-reaction of the quantum fields on the bulk geometry.  But expansions of both the area and $S_{QFT}$ near the splitting surface of such back-reacted saddles must nevertheless exist, and they will satisfy the same replica symmetry and compatibility with Euclidean expansions described above.  So again both must be extremal at $\fixM$ in saddles preserving both replica and conjugation symmetry.

\section{Discussion}
\label{sec:disc}

Our work above studied saddle points of real-time gravitational path integrals associated with the (annealed swap) R\'enyi entropy $S_n(D)$ for domains of dependence $D$ on some asymptotically anti-de Sitter boundary for integer $n>1$.   For simplicity we considered pure Einstein-Hilbert gravity (in any dimension), but many other cases follow by dimensional reduction.  In addition, footnote \ref{foot:HD} describes how it may be generalized to gravity theories with perturbative higher derivative corrections by making use of further results from \cite{Dong:2019piw}.   So long as the full system satisfies an appropriate analogue of the GSL, extending the arguments to any theory of matter governed by a local two-derivative action with perturbative higher-derivative terms appears to be merely a technical exercise.

We first worked at the level of the leading order terms in the stationary phase approximation, but we then included quantum corrections to all orders.  We explicitly assumed our theory to satisfy the quantum focussing condition.   When the saddle preserves replica and conjugation symmetries, we then showed under a certain technical assumption  that the splitting surface $\fixM$ to be causally inaccessible from $D$.  As a result, we could deform the saddle without changing $S_n(D)$  to make manifest that $S_n(D)$ is independent of both the choice of any sources on $D$ and the choice of any Cauchy surface $\Sigma_D$ for $D$.  One may thus say that (annealed swap) R\'enyi entropies are time-independent in the sense associated with unitary quantum theories living on the asymptotic boundary.  However, this argument involved a technical assumption (see the end of section \ref{sec:saddles}) about solutions to the gravitational initial value problem for which the induced metric on the initial surface may fail to be positive definite.  This assumption deserves to be better understood.

Recall then that unitarity is a key property of quantum mechanics, and that the study of quantum gravity has long sought to understand whether and in what sense unitarity might hold.  In particular, in a baby universe scenario one should distinguish between unitary evolution of the full quantum gravity Hilbert space and unitarity `from the perspective of an asymptotic observer,' by which we mean unitarity on each superselection sector for the algebra of asymptotic observables.

Now, it is natural for the above notions of unitarity to be closely related in a theory of quantum gravity.  After all, one expects the gravitational Hamiltonian to be a boundary term, and thus to lie in any algebra of asymptotic observables.  But this then immediately implies that it preserves the associated superselection sectors \cite{Marolf:2008mf,Marolf:2008tx}.

In particular, a full proof of this unitarity follows if one adopts the axiomatic framework for gravitational path integrals described in \cite{Marolf:2020xie}.  The axioms of that reference are stated in terms of Euclidean path integrals, so we should in fact add the additional axiom that Lorentzian time-evolution is given by a Wick rotation of that framework.  Under such assumptions, section 4.1 of \cite{Marolf:2020xie} shows that the gravitational Hamiltonian is self-adjoint and lives in the algebra of boundary observables, so that $e^{iHt}$ is unitary and preserves superselection sectors.  In fact, section 4.1 of \cite{Marolf:2020xie} also shows the density of states in each superselection sector to be bounded by $e^{S_{BH}}$.  So then BH unitarity holds whenever this bound is saturated.

On the other hand, the current understanding of quantum gravity path integrals is sufficiently poor that such formal arguments are naturally regarded with suspicion.  Furthermore, the conformal factor problem of Euclidean gravity \cite{Gibbons:1976ue,Gibbons:1978ji} then amplifies such concerns when Euclidean path integrals appear to play a fundamental role (though there are good reasons to suspect that this is not a serious issue in the end \cite{Hartle:1988xv,Dasgupta:2001ue,Anninos:2012ft,Cotler:2019nbi,Benjamin:2020mfz}).  As a result, more concrete tests of unitarity --  such as the derivations of the Page curve and the tests described here -- provide important pieces of evidence that the above formal arguments are physically meaningful.  It thus remains of great interest to move beyond the limitations of the current work to address more general situations.

One extension turns out to be straightforward.  This is the generalization to the case of non-integer replica numbers $n>1$.  At some level, time-independence of R\'enyi entropies for non-integer $n$ must follow by analytic continuation from the integer $n$ result derived here.  But one may also give a more direct argument using the Lewkowycz-Maldacena trick of describing a replica-invariant saddle ${\cal M}_n$ by its ${\mathbb Z}_n$ quotient $\tilde {\cal M}_n  = {\cal M}_n/{\mathbb Z}_n$.  In terms of $\tilde {\cal M}_n  = {\cal M}_n/{\mathbb Z}_n$, the boundary conditions for $\tilde {\cal M}_n$ do not depend on $n$, but $\tilde {\cal M}_n$ has a conical singularity whose strength does depend on $n$. While Lewkowycz and Maldacena worked in Euclidean signature, the analogous trick can also be used directly in Lorentz signature using the associated notion of `conical singularity' (see e.g. \cite{Colin-Ellerin:2020mva}).  This description is then straightforward to analytically continue to non-integer $n$.  Furthermore, all of the power series expansions used in our work continue to hold on such $\tilde {\cal M}_n$ in the obvious way.  For $2>n>1$ one finds that the Riemann tensor can be singular at $\tilde X^\pm=0$ but, for vacuum Einstein-Hilbert gravity with cosmological constant, the equations of motion require the Ricci tensor to be proportional to the metric.  In particular, $R_{ab}k^ak^b=0$ for any null vector $k^a$, so one may continue to use the Raychaudhuri equation as in \cite{Wall:2012uf,Headrick:2014cta}. With this understanding
one may repeat our arguments verbatim for non-integer $n>1$.  Once again, the conclusion is that $S_n(D)$ depends only on $D$ and not on the choice of a particular Cauchy surface $\Sigma_D \subset D$.

On the other hand, new input will clearly be required to generalize the symmetry-based arguments of this work to address saddles in which replica symmetry is broken.  And while replica-invariant saddles appear to dominate in many situations, it has recently been shown \cite{Penington:2019kki,Dong:2020iod,Marolf:2020vsi,Akers:2020pmf} that replica-breaking saddles have important effects near HRT phase-transitions. Furthermore, even a small sub-dominant effect that violates unitarity would be of great interest.  We will therefore return to the question of replica-breaking saddles in a forthcoming work.

\acknowledgments
It is a pleasure to thank Xi Dong, Henry Maxfield, Pratik Rath, and Douglas Stanford for many useful discussions.  This work was supported by NSF grant PHY1801805 and funds from the University of California.

\appendix
\section{Replica symmetry and the entropy of a holographic theory on ${\cal M}_n$: an example}
\label{app:RSHRT}

This appendix provides a simple example illustrating the discussion in section \ref{sec:holographic} of $S_{QFT}$ for holographic field theories on ${\cal M}_n$.  For simplicity, we consider Jackiw-Teitelboim gravity with two boundaries, and we focus on the Renyi problem associated with the entropy of a single boundary in the vacuum state.  The classical real-time replica wormholes ${\cal M}_n$ for this problem were constructed in \cite{Colin-Ellerin:2021jev}, which found that they may be built from $2n$ manifolds $M_{-}^i, \tilde M_-^j$  each with the identical metric
\begin{equation}
\label{eq:JT}
ds^2=\frac{4(\tilde{x}^+ \tilde{x}^-)^{\frac{1}{n}-1}d\tilde{x}^+ d\tilde{x}^-}{n^2(1-(\tilde{x}^+ \tilde{x}^-)^{\frac{1}{n}})^2}.
\end{equation}
 The AdS boundaries are located at $\tilde{x}^+ \tilde{x}^-=1$.
 We take the splitting surface $\fixM$ to lie at $\tilde{x}^+ =\tilde{x}^- =0$ and we sew the replicas together in the usual way along arbitrary spacelike surfaces connecting $\fixM$ to the right AdS boundary (where $\tilde x^\pm >0$).

We now couple our JT-gravity system to a holographic quantum field. To make the example non-trivial, we will allow the quantum field to be in any member of a one-parameter family of states labelled by an amplitude $A$, with $A=0$ being the vacuum state.  In general, we would expect there to be some back-reaction on the metric \eqref{eq:JT} for $A \neq0$.  But for simplicity we will choose a model for which this back-reaction vanishes at the order where $S_{QFT}$ is given by the classical HRT entropy.

In particular, we take our holographic QFT to be dual to gravity on an AdS$_3$ spacetime with boundary metric given by \eqref{eq:JT}.  Since the boundary metric itself has a boundary, we also take the bulk to be truncated by a dynamical EOW brane anchored to the AdS boundaries of \eqref{eq:JT}.

As described in section \ref{sec:holographic}, the relevant HRT surfaces $\gamma_{\cal N}$ will stretch from some point $\gamma$ in the 2-d JT-gravity spacetime \eqref{eq:JT} to this end-of-the world brane.

It will be convenient to define our model so that the bulk spacetime ${\cal N}$ is always described by the same bulk metric and such that it always has vanishing boundary stress tensor.  We can do so while still allowing non-trivial fluctuations in the bulk HRT entropy by including a dynamical field $\phi$ that lives on the EOW brane.  We take this $\phi$ to be a massless scalar on the brane so that the bulk action takes the form
\begin{equation}
\begin{aligned}
I=-\frac{1}{16 \pi G} \int_{\mathcal{N}_n} d^3x(R-2\Lambda)-\frac{1}{8 \pi G} \int_{EOW} d^2x \sqrt{h} K -\int_{EOW} d^2x \sqrt{h} h^{ij} \partial_i \phi \partial_j \phi
\end{aligned}
\end{equation}
with appropriate counterterms. Here $h_{ij}$ is the induced metric on the EOW brane.

Our one-parameter family of states will then be defined by imposing the boundary condition
\begin{equation}
\phi_b=A \left( \tilde{x}^+ +\tilde{x}^-\right)
\end{equation}
on every $M_-^i$, $\tilde M_-^i$. As we can see, this boundary condition has the $\mathbb{Z}_n$ replica symmetry.
We now solve for the backreaction caused by the EOW-brane scalar field.
The equation of motion for $h_{ij}$ gives
\begin{equation}
K_{ij}-h_{ij}K=8\pi G T_{ij}
\end{equation}
where
\begin{equation}
T_{ij}=\frac{2}{\sqrt{h}}\frac{\delta I_{\phi}}{\delta h^{ij}} =  \partial_i \phi \partial_j \phi - \frac{1}{2} h_{ij} h^{kl}  \partial_k \phi \partial_l \phi.
\end{equation}
The equation of motion for the scalar field gives
\begin{equation}
\partial_i (\sqrt{h} h^{ij}\partial_j \phi)=0
\end{equation}

The boundary metric \eqref{eq:JT} suggests that we describe the bulk using the hyperbolic slicing of AdS$_3$ and replace each hyperbolic slice with the $R=-2$ metric \eqref{eq:JT} so that the bulk metric takes the form
\begin{equation}
\label{eq:dhbulk}
ds^2=d\rho^2+\cosh^2\rho \left( \frac{4(\tilde{x}^+ \tilde{x}^-)^{\frac{1}{n}-1}d\tilde{x}^+ d\tilde{x}^-}{n^2(1-(\tilde{x}^+ \tilde{x}^-)^{\frac{1}{n}})^2}\right).
\end{equation}
Without an EOW brane the coordinate $\rho$ would range over the entire real line:  $-\infty<\rho<\infty$.  However, we take the boundary metric \eqref{eq:JT} to live at $\rho=-\infty$, and the EOW  brane will cut off the bulk at some $\rho = \rho_{EOW}(\tilde x^+, \tilde x^-)$. When no matter field is present, the brane lies at $\rho_{EOW}=0$, where the extrinsic curvature vanishes.

We wish to solve the equations of motion to find $\rho_{EOW}$ perturbatively in $A$. At zeroth order we have $\rho=0$ on the brane.  We may thus choose boundary conditions in the far past (or on a Euclidean piece of an appropriate Schwinger-Keldysh contour) so that at this order we have
\begin{equation}
\phi = \phi_0 :=A \left(\tilde{x}^+ +\tilde{x}^-\right),
\end{equation}
Its stress tensor on the brane is $T_{++}=T_{--}=A^2$, and the trace vanishes, $T_{+-}=0$.

To compute backreaction, we take the brane to be located at $\rho= \rho_1(\tilde{x}^+,\tilde{x}^-)$, whose normal is
\begin{equation}
n_\mu dx^{\mu}= \frac{1}{N}\left( d\rho+ \partial_+ \rho_1 d\tilde{x}^+ +\partial_- \rho_1 d\tilde{x}^- \right).
\end{equation}
To leading order in $\rho_1$, the EOW brane extrinsic curvature $K_{\mu \nu}=\nabla_\mu n_\nu$ has components
\begin{equation}
K_{++}=\frac{1}{n \tilde{x}^+} \left(\frac{2}{1-(\tilde{x}^+ \tilde{x}^-)^{\frac{1}{n}}}-n-1 \right) \partial_+ \rho_1 -\partial_+^2 \rho_1
\end{equation}
\begin{equation}
K_{--}=\frac{1}{n \tilde{x}^-} \left(\frac{2}{1-(\tilde{x}^+ \tilde{x}^-)^{\frac{1}{n}}}-n-1 \right) \partial_- \rho_1 -\partial_-^2 \rho_1
\end{equation}
\begin{equation}
K_{+-}=-\frac{2(\tilde{x}^+ \tilde{x}^-)^{\frac{1}{n}-1}}{n^2\left(1-(\tilde{x}^+ \tilde{x}^-)^{\frac{1}{n}}\right)^2}\rho_1-\partial_+ \partial_- \rho_1
\end{equation}

Solving the equations of motion
\begin{equation}
K_{ij}=8\pi G T_{ij},
\end{equation}
one finds that the position of the EOW brane is
\begin{equation}
\rho_1(\tilde{x}^+,\tilde{x}^-)=-\frac{8\pi G A^2 n}{4n^2-1} \frac{ (1+2n-(2n-1)(\tilde{x}^+ \tilde{x}^-)^{\frac{1}{n}})}{2(1-(\tilde{x}^+ \tilde{x}^-)^{\frac{1}{n}})} \left( (\tilde{x}^+)^2+(\tilde{x}^-)^2\right).
\end{equation}

The entanglement entropy of $\Sigma_\gamma$ is given by the length of the HRT surface (geodesic) that goes from $\gamma$ to the EOW brane. For our choice of coordinates, the geodesics of interest are given by $(\tilde{x}^+,\tilde{x}^-)=const$, $-\infty<\rho<\rho_1$.  Thus the entanglement entropy of $\Sigma_\gamma$ is given by
\begin{equation}
S_{QFT}[\gamma]=\frac{1}{4 G_{N}} \int_{-1/\epsilon}^{\rho_1(\tilde{x}^+,\tilde{x}^-)} d \rho=\frac{\rho_1(\tilde{x}^+,\tilde{x}^-)+1/\epsilon}{4 G_{N}}.
\end{equation}
where we have introduced a cutoff at $\rho=-1/\epsilon$.
Rewriting the above formulae in terms of the coordinates $\tilde X^{\pm}\equiv (\tilde{x}^\pm)^\frac{1}{n}$, it is then easy to see that first derivatives of $S_{QFT}$ with respect to $\tilde X^\pm$ vanish at the splitting surface $\fixM$ ($\tilde{x}^+=\tilde{x}^-=0$):
\begin{equation}
\frac{dS_{QFT}[\fixM]}{d \tilde X^{\pm}} =0.
\end{equation}

\section{The shadow as a Wick rotation}
\label{app:figs}

This appendix illustrates how, when the real-time saddle ${\cal M}_n$ is a Wick rotation of some ${\cal M}_n^E$, the extended shadow $\hat {\cal S}_R$ can be taken to be defined by a different (and smoother!) Wick rotation of ${\cal M}_n^E$.

\begin{figure}
\centering
\includegraphics[width=\linewidth]{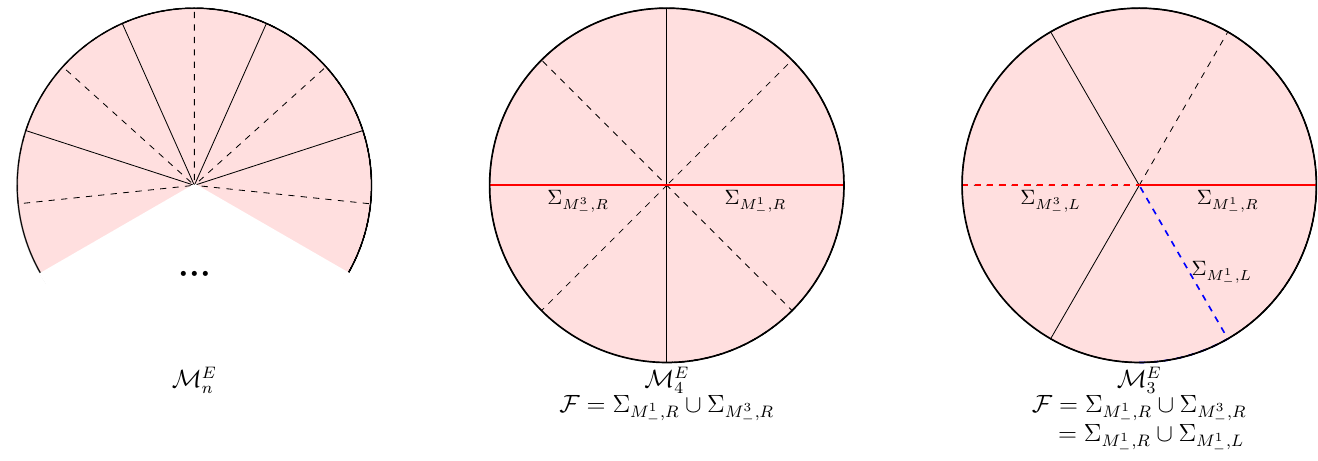}
\caption{{\bf All panels:} Examples for various $n$ of the Euclidean manifolds $\mathcal{M}_n^E$ that can be Wick rotated to obtain the real-time saddle $\mathcal{M}_n$.  We take these to have replica and conjugation symmetries. Each conjugation symmetry leaves invariant some combination of dashed and solid black surfaces.  As a result, the induced metric is real on these surfaces and the Euclidean-signature extrinsic curvature is imaginary.  Thus Wick rotation to Lorentz signature gives real Cauchy data on these surfaces. {\bf Center and Right:}
For the cases $n=4$ and $n=3$, we may consider symmetries that act simultaneously by complex conjugation and by reflection across the red surfaces $\mathcal{F}$ shown.
In each case, we may take $\mathcal{F}$ to be the surface $\tau=0$ and then Wick rotate $\tau$ to define a Lorentz signature spacetime.  The result is not $\mathcal{M}_n$ (which is given by a different Wick rotation).  Instead, it gives a valid extended right shadow $\hat {\cal S}_R$.  In the even case (center), this shadow $\hat {\cal S}_R$ has a right/left ${\mathbb Z}_2$ reflection symmetry that for swaps the isometric surfaces $\Sigma_{M^{n/2+1}_-,R}$ and $\Sigma_{M^{1}_-,R}$.  But in the odd case the fact that all dashed surfaces are related by replica symmetry means that  $\hat {\cal S}_R$ is a smooth manifold whose initial data on $\mathcal{F}$ matches that of any $M_-^i$ on $\Sigma_{M_-^i}$. }
\label{fig:ME}
\end{figure}

As mentioned in footnote \ref{foot:shadow}, any such
${\cal M}_n^E$ must have a ${\mathbb Z}_2$ symmetry that acts simultaneously by a reflection across a fixed-point set ${\cal F}$ and by complex-conjugating all sources.  This symmetry is a consequence of hermiticity for each copy of the density matrix employed in the replica trick.  The set ${\cal F}$ thus partitions ${\cal M}_n^E$ pieces, each containing $n/2$ replicas (whether $n$ is odd or even).  The extended shadow $\hat{\cal S}_R$ can then be defined by introducing a Euclidean time $\tau$ for which ${\cal F}$ is $\tau=0$ and then performing the standard Wick rotation in terms of $\tau$.  For even $n$, both the of ${\cal F}$ to the left and right of $\fixM$ coincide with surfaces $\Sigma_{M_-^i,R},\Sigma_{M_-^j,R}$ so that $\hat{\cal S}_R$ has a ${\mathbb Z}_2$ reflection symmetry, while for odd $n$ the left part of ${\cal F}$ instead coincides with some $\Sigma_{M_-^j,L}$ (which by replica symmetry has the same geometry as  $\Sigma_{M_-^i,L}$). See figure \ref{fig:ME}.


\bibliographystyle{JHEP}
\bibliography{unitarity}

\end{document}